\documentclass[11pt,a4paper]{article}
\usepackage[utf8]{inputenc}
\usepackage{amsmath}
\usepackage{amsfonts}
\usepackage{amssymb}
\usepackage[english]{babel}

\usepackage{color}
\usepackage{epsfig}
\usepackage{cite}
\usepackage{graphicx}%
\usepackage{dcolumn}%
\usepackage{bm}%

\begin{document}
\begin{center}
{\large Parameter space stability of multiple soft interactions} \\
\vspace*{0.5cm}
Rahul Kumar Thakur\textsuperscript{a,}\footnote{thakurr58@gmail.com}, Bhupendra Nath Tiwari\textsuperscript{b,c,}\footnote{bhupendray2.tiwari.phd@iitkalumni.org}, and Rahul Nigam\textsuperscript{a,}\footnote{rahul.nigam@hyderabad.bits-pilani.ac.in}\\
\vspace*{0.5cm}
$^a$ BITS Pilani, Hyderabad Campus, Hyderabad, India.\\
$^b$ INFN-Laboratori Nazionali di Frascati,	Via E. Fermi 40, Frascati 00044, Rome, Italy.\\
$^c$ University of Information Science and Technology, ``St. Paul the Apostle", Partizanska Str. bb 6000 Ohrid, Republic of Macedonia.\\
\vspace*{0.1cm}
\end{center}
\begin {abstract}
We study the fluctuation theory analysis of multiple soft interactions that are used to design SIBYLL 2.1 and related cosmic ray event generators. We examine fluctuation stabilities of Lund string fragmentation involving quark-antiquark and diquark-antidiquark pairs. Using the primordial transverse momentum hadron-hadron pairs, we find that the resulting Gaussian distribution yields an ill-defined ensemble under fluctuation of the model parameters of parent quarks and diquarks. Further, we investigate the nature of multiple soft interactions through Pomeron-Reggeon fluctuations at various energy scales. We show that there are both stable and unstable regions under fluctuations of the Reggeon and Pomeron densities at all initial energy scales. We find that the optimal stability zone is achieved at the Reggeon and Pomeron densities as predicted by GRV model. For hadron-hadron collisions, the limiting Gaussian profile as a function of the impact parameters of soft interactions for proton-proton collisions corresponds to an indeterminate statistical basis. Subsequently, we add higher order corrections in the Regge trajectory of constituent hadrons that play an important role in determining the shape of a proton or meson profile function in the realm of minijet model. We also discuss the stability of accelerated nucleons with jerks and higher order contributions as well as soft profile functions at off-shell-conditions. In addition, we illustrate the wall of stability/ instability for multiple soft interactions by considering energy dependent soft contributions to scattering cross section as the model embedding. Qualitative discussions are provided at physical energy scales ranging from $1$ GeV to $1800$ GeV. This includes the analysis of CDF, P238, UA5 and ZEUS experiments
\end{abstract}
\newpage

\section{Introduction} 
QCD finds various applications to understand the behavior of quarks that we apply in examining astroparticle physics and the optimal detection of astroparticles \cite{astro} from deep space. Cosmic ray event generators \cite{SIBYLL} are understood through extensive air showers \cite{EAS}, Monte Carlo simulations \cite{MC}, dual parton model (DPM) and Lund fragmentation of strings \cite{SIBYLL} that connect two quarks or diquarks. In particular, DPM picture allows ones to understand a nucleon as the composition of quarks and diquarks. Here, the quark as a color triplet and diquark as a color antitriplet are mediated via soft gluonic interactions \cite{SIBYLL}. Physically, it is worth mentioning that a projectile quark or diquark gets combined which a target diquark or quark, whereby they form two strings \cite{6b}. Because of a high tension, it is well-known that such strings get fragmented and they obey the Lund string fragmentation \cite{SIBYLL, 6b}. 

The aim of this paper is to offer a statistical stability analysis of hadronic interactions and their associated phase transition phenomena satisfying the Bjorken scaling and low transverse momentum of the scattered partons. In this regard, one has a steeply increasing gluon density at fractional energy as exhibited in the results of our paper \cite{Ourgluonpaper}. The motivation of this paper is to determine criteria for the stability of a gluonic configuration under the variations of its parameters. This support to clarify the regions where hadronic decays happen. In brief, we determine the parameter space domains where new particles are pron to be produced or the considered ensemble of particles remains stable during the collision processes of hadrons. Further, we provide qualitative discussion and future directions for the study of random observation samples with multiple soft interactions. 
	
Specific focus is given to examine the nature of decays and instabilities in hadronic systems. At various energy scales, we have offered detailed analysis of phase transitions corresponding to particular processes, and collaborations including CDF, P238, UA5, and ZEUS that take place at a given energy scale. The respective numerical values are depicted in Tables 1-7. Hereby, we have shown that the predicted values arising from our model are well compatible with their phenomenological counterparts that arise in the case of SIBYLL \cite{SIBYLL}.

In this paper, firstly, we study fluctuation theory analysis of Lund string fragmentation by considering quark or diquark distribution profile in the space of model parameters of probability distribution, see \cite{fta} towards effective cross section and LHC event data. In particular, there are recent developments towards the understanding of the phase structure, quantum phase transitions and particle production in the realm of nuclear collisions \cite{LHC}. With this motivation, we treat the distribution profile as a probability distribution in the space of effective quark mass and associated distribution indices. This allows studying stability of hadronic models under fluctuations of their parameters. Our treatment follows from the fact that the fragmentation of strings happens by generating a quark-antiquark pair or diquark-antidiquark pair that are attached to the ends of QCD strings \cite{SIBYLL}. In this setting, we focus on the fractional energy $x$ and the centre of mass energy square $s$ of hadron-hadron interactions at which the collision occurs. In order to have a normalized probability distribution, $x$ is kept as an arbitrary number between $(0,1)$ and $s$ as a finite real number.

Hadrons are formed by combining new flavors with the existing ones. In the primordial model, when quark-antiquark pairs have equal magnitude and the opposite direction transverse momentum, the quark distribution profile turns out to be a Gaussian distribution that depends on the mean energy and center of the mass energy of hadron-hadron interactions \cite{hhi}. In this case, we have shown that the discriminant considering the Gaussian distribution profile as the model embedding, yields a degenerate statistical system for all values of the model parameters. Note that there is a threshold mass \cite{SIBYLL} up to which the fragmentation of strings can continue. For the above cases, consider quark or diquark configuration with a threshold mass that is typically in the range of 0.9 GeV to 1.3 GeV, whereby two final hadrons are formed. In this system, as far as the fluctuation theory analysis is concerned, the fraction of energy between the new particle and parent quark or diquark, the transverse mass is kept arbitrary in order for the happening of a reaction. 

In this regard, we focus on minijet model in order to study QCD-improved parton models. Here, we have provided fluctuation theory analysis of gluon density in the limit of double leading logarithmic approximation \cite{Ourgluonpaper}. Notice that our analysis is capable of characterizing the matter stability as well as formation of new particles. Further, we take an account of the high energy fluctuations \cite{24} and geometric saturation condition \cite{SIBYLL} that is obeyed by the strong coupling constant and gluon density in the framework of effective field theories. Our analysis goes well beyond the outcomes of SIBYLL 1.7 by having a energy dependent transverse momentum cutoff as in the standard minijet model. As in the Feynman diagrams, this is because of the fact that the production of $n$ pairs of minijets has it cross section that is exactly $n$ cut parton ladders that are summed up over all uncut ladders, see \cite{28b} for an overview of multiplicity distributions of highly energetic particles and associated dual partons.

Following the above analysis we concentrate on multiple soft interaction inspired by Regge theory \cite{30b}. Namely, it is observed that SIBYLL 1.7 is too simple with an energy independent soft contribution to the profile function in Gribov-Regge (GR) theory \cite{Drescher}. This is very similar to hard interactions due to a low energy inelastic collision with the cross section of about $32$ mb. Due to this drawback, SIBYLL 2.1 \cite{SIBYLL} was introduced in order to cover a large portion of phase space to properly understand the soft interaction physics by allowing multiple soft interactions \cite{msi}. This is motivated by Regge theory \cite{30b} where the soft cross section is defined as a sum of two power laws where the first one is for Pomeronic interactions and the other one is for the Reggeonic interactions \cite{SIBYLL} as their weighted contributions. In this case, we exploit fluctuation theory analysis \cite{bntef, bntfr} over space of model parameters. Here, we focus on the scattering cross section as the model embedding function in the space of Reggeon and Pomeron densities and the centre of mass energy square, starting from the multiple minijet productions. 

We consider uncertainties in the transverse momentum of partons with respect to their spreads over the region of soft interactions, see \cite{bntun} for general form of uncertainty relations at small length scales. We offer special attention to Gaussian soft profile function that has an energy dependent variance over the region of interactions. In practice, this involves the leading value of the variance over the impact parameter of soft profile as the transverse size of a proton in its low energy limit. Under the variations of model parameters, we examine the fluctuations of protons and meson profiles \cite{meson}. Our analysis shows that the concerned probability distribution is well capable of generating new events, whereby it is shown that soft interactions \cite{soft} are mediated through gluons. It is worth recalling that there is a gluon density saturation \cite{SIBYLL} that happens around $1$ GeV, whereby the soft mass is required for strings and quark sea in order to have regulated singular part of the distribution profile that is essential for Lund string fragmentation. Motivations are provided in the light of diffraction dissociation events that happens to achieve the final state with a large rapidity gap, see \cite{SIBYLL} for an associated discussion. 

Hereby, our fluctuation theory model supports understanding of diffraction physics and QCD phenomenology. Such studies are well motivated by Pomeron hadron interactions where the final states are described by multiple soft and hard interactions, .e.g., $\pi-\pi$ interaction. Indeed, there are experimental motivations from data science, for instance, see UA4 collaboration \cite{37b}. This offers an improved understanding of hadron-nucleon and nucleon-nucleon interactions. Our analysis incorporate elastic and quasi-elastic interactions without the production of new particles \cite{11b}. String fragmentation \cite{6b} fluctuations are equally applicable to partonic system where the string fragmentation happens for color connected partons. As extensions of our model, future directions include stability analysis of nucleus-nucleus interactions, high energy air showers, collider experiments, fixed target experiments \cite{SIBYLL} inclusive of the cross section for charged particle, pseudo rapidity analysis and the optimal determination of central region of the soft interactions that take place in a given reaction.

In the light of hadronic interactions, we focus on the stability of configurations involving string frangmentations and multiple soft interactions. Our model is consistent with previous measurements such as SIBYLL 1.7 and its updated version SIBYLL 2.1 \cite{SIBYLL}. As far as the SIBYLL model based event generators are concerned, on the one hand, our analysis allows to interpret new set of data from collision experiments as the snapshots of what a proton really looks like at a given instant \cite{1}. On the other hand, for hard processes such as the Higgs boson or events mediated via certain high transverse momentum jets, the associated rates and event generation properties are examined by using perturbation theory techniques \cite{14}. Hereby, the hard interactions happen at a point that happens as approximately in the cases when we have a constant value of the transverse momentum of the QCD improved partons \cite{23}. 

Our proposal offers an apt understanding of particle productions and associated stability analysis of partons. In particular, we consider the quark or diquark distribution function as the embedding profile, whereby we discuss its saddle point structures and stability criteria. One may extend the above analysis to compute the same for soft terms to examine the stability of the above hadronic interactions under variations of the model parameters. Moreover, it is instructive to note that our analysis provides the nature of soft interactions under fluctuations of Reggeon and Pomeron densities as in \cite{SIBYLL}. It would be worth computing the torsion of multiple soft interactions, see \cite{Thomas} for an overview. Such investigation are left open for future research and developments. 

Further improvements exist involving the energy dependence, QCD phenomenology and the transverse momentum of colliding particles. Hereby, our analysis provides the nature of phase transition points. It is noted that a complete analysis can be provided by considering the full Glauber model of nucleus-nucleus interactions \cite{SIBYLL}. Such investigations will be developed in the future work. Possible directions include nucleus-nucleus interactions, high energy air showers, analysis of collider experiments, fixed target experiments inclusive of the cross section for charged particle pseudo rapidity analysis and the optimal determination of central region of soft interactions. This is well supported through our stability analysis of various hadronic systems and their experimental counterparts involving multiple soft interactions in relation to dual parton model and QCD-improved partons.

Finally, it will be equally interesting to extend our analysis for strange particle production such as $\Lambda^0$ and $K_s^0$ and the analysis of large rapidity fluctuations in the light of astroparticles and their production in deep sky \cite{SIBYLL}. This idea will be extended in relation to $u$, $d$, $s$ quarks, diquarks and gluons in order to offer the optimal designing of detectors to observe muons, astrophysical particles and neutrinos. In this regard, we provide a brief overview of the quark density function in the framework of the fluctuation theory. The concerning discussion is relegated to section 2 below. With this motivation, we explore the stability structures of hadronic systems with multiple soft interactions in the light of background color fluctuations in minijet models and particle collisions \cite{Wang}. Namely, we analyze elementary particle density fluctuations to understand extensive air showers of cosmic rays. Prospective study of its next version and a comparative analysis of LHC data \cite{LHC} are left open for future research. 

The rest of the paper is organized as follows. In section 2, we provide a brief review of the model towards the understanding of statistical stability. In section 3, we overview the nature of hadron density in the designing of cosmic ray event generators. In section 4, we give stability properties of partons in DPM picture and analyze in limiting Gaussian systems. In section 5, we offer discussion of the results and their comparison with existing models. In section 6, we extend our analysis for collisions with translated soft interaction region and inclusion of arbitrary higher order corrections in the Regge trajectory of constituent hadrons with a varying boundary of soft interactions. Finally, in section 7, we conclude the paper with prospective directions for future research and developments.    
\section{Physics at High Energies: An Overview up to $1$ TeV}
In this section, we provide an overview of the hadron density fluctuations in the light of cosmic ray event generators. It is worth recalling that hadronic reactions provide a way to study the properties of the nuclear matter and their constituents. Thereby, in order to allow multiple soft interactions as in SIBYLL 2.1 , the eikonal for soft interactions is described using Regge theory, thus, our focus lies on the statistical stability of hadrons at different energy scales, e.g., the collision of two protons in heavy nuclei at particle colliders such as the Relativistic Heavy Ion Collider (RHIC) at Brookhaven National Laboratory \cite{rhic} and Large Hadron Collider (LHC) at the CERN laboratory \cite{cern}.%

In particular, we focus on the physical observables, the optimal designing of cosmic ray event generators, photo nuclear processes, ultra high-energy neutrinos and their stability perspectives through fluctuations of quarks, diquarks, gluons and plasma states. In the light of SIBYLL 2.1 event generators, this provides an intrinsic characterization of soft interactions and diffraction dissociation reactions.
\subsection{Cosmic rays}
First of all, as far as the observables are concerned, our universe passes a large extent that has a few trillion galaxies and scatters across the depth of the space 
\cite{cosmology,cosmology1,cosmology2}. For a chosen system, there exists new measurements concerning the cosmic ray event generators. Here, in the realm of hadronic physics, one uses a soft type of interactions than the previously known endeavors \cite{SIBYLL}. The revised measurement coming from multiple soft interactions lies at the center of the present research. This gives a new interpretation to the fundamental properties of ultra high energy physics. 

Hereby, the exact rate at which the Reggeon-Pomeron configuration is getting into the equilibrium depends on the value of the initial energy scale. Till date, it is a highly elusive matter that is an important subject of modern high energy research. 
In relation to the cosmic rays and  the ultra high-energy (UHE) neutrinos towards the designing of associated event generators, it is worth mentioning that various ultra-high-energy neutrinos at their energies $> 1$ GeV may be produced by certain photo nuclear processes. This involves UHE cosmic rays and photons falling either at a source or during their propagation involving a cosmic microwave background radiation \cite{1, 2}. 

An expected low density flux \cite{3} having a small cross-section \cite{4} implies a vast detection volume that is needed towards a successful detection of cosmic rays and photons. There are mechanisms that offer a coherent increase of Cerenkov emission at long wavelengths \cite{5}. This results via extra negative charges that are produced in particle showers in a dense media. Along with a long  radio wave length say for instance of a few km, we may use of abundantly available Antarctic ice layers towards the detection of UHE neutrinos, see \cite{6} for a detailed introduction.

\subsection{Quarks, gluons and plasma states}
In various collision of elementary particles, for example, quarks, gluons or  electrons and positrons, jets are created in their hard collisions. The leading jet particles go out from the collision zone with a large transverse momentum and they fragment into a nearly collimated flow of hadrons termed as the jet cone. Such jets are well identified and these processes are studied both theoretically and experimentally \cite{SIBYLL}. Having embedded the hard collisions as in the environment, the situation for the high energy heavy ion collisions is more complicated than we intend to address in future.

The leading jet particle could interact with both the environment and the plasma of quarks and gluons that lead to a transfer of energy/ momentum to the medium. In this concern, the physics of quarks is expanding its applications through the formation of the QGP state. Namely, in a medium, the fragmentation of jet particles into hadrons gets modified. In a low energy regime, hadrons have a velocity less than or equal to the QGP expansion velocity. Both these processes change through the distribution of hadrons, whereby the jet cones are intensively studied both theoretically and experimentally. Largely, the hadrons as created in the above heavy ion collisions arise as a phase of the QGP. Their multiplicity is described by the statistical model, whereby we offer the related calculations from the perspectives of the statistical fluctuation theory. 

Therefore, such particles have no information concerning the interaction of partons. Probabilistically, jet partons show nonequilibrium behavior with the partons esteeming from the QGP. The hadrons thus go through jet fragmentation that may be among the viable sources, offering information about the underlying parton-parton interactions \cite{11b}. In short, the present study arises an important objects concerning the above high energy configurations.

Following the developments at RHIC and LHC, we focus on the heavy ion physics, whereby we study QCD matter structures in the limi of high energy density and temperature. This supports studying the underlying dynamics of jet particle production and medium interactions as a multi-stage phenomenon. As far as the physical understanding of heavy-ion collisions is concerned, hard interactions govern elementary scattering processes and subsequent branching that scale down to certain non-perturbative scales. Such phenomena happen both in the vacuum as well as in media. Soft scales arise at the order of the temperature of a given medium. 

In the perspective of LHC physics, we explore interactions of soft partons as produced in air showers that is strongly coupled with the medium. Soft scales govern the hadronization processes that takes place in the vacuum with some sufficiently energetic probes. Discussion of different processes involving the jet shower evolution in a medium as well as the designing of onset scales are among the important problems. The relation between the above processes leads to modifications in the longitudinal and transverse distributions of constituent jet particles with respect to their vacuum fragmentation. In practice, such  modifications support  investigating jet shape observable that constrain the dynamics of  energy loss in a given medium through the color coherence and medium properties \cite{37b}. These can be mediated through the temperature fluctuations and the evolution of the medium degrees of freedom that is the density with a definite resolution scale \cite{14}. Energetic quarks and gluons including partons fragment, whereby they produce collimated bunches of hadrons that are termed as the jets of hadrons. 

Physically, jets conserve both the energy and direction of the emanating partons. Consequently, the jets are in use at colliders for generic partons. Indeed, such applications are under use at CERN Large Hadron Collider (LHC). In the recent years, various extensive developments that are well beyond the basic use, for example, we could understand the underlying physical setup if the parton is a quark or a gluon, that is, by identifying the rare cases where a single jet originates from a sequence of multiple hard partons. Such events follows from the rules of hadronic decays involving high energy particles such as $W$, $Z$ or Higgs bosons, and top quark or other massive objects \cite{16,17,18, 19}.  

At high energies, the jet substructures as developed above support exploiting the full kinematics as discussed in the framework of the LHC. Notably, the high transverse-momentum that is, a high $p_T$, region help to maximize the LHC's sensitivity to hadronic scenarios of new physical manifestations \cite{20}. High energetic partons that are usually produced in high-energy hadron-hadron collisions initiate a cascade of events involving various lower-energy quarks and gluons. Such particles hadronize into collimated sprays of colorless hadrons \cite{SIBYLL}. This gives an alternative definition of jets.

Monte Carlo event generators of particle collisions, viz. PYTHIA \cite{21b}, describe both the perturbative cascade and dominated cascades through soft gluon emissions. In a specific case, we have collinear parton splittings, where the final hadronizations are described by non-perturbative models. This captures the physics at the end of the parton shower that happens below a cutoff of the order of $1$ GeV. In heavy ion collisions and events that happen at LHC and at RHIC, the hadronic high transverse momentum observables get deviated from certain baseline measurements in proton-proton collisions. These models are referred as the jet-quenching. It was established at the RHIC at a single inclusive hadron spectrum and high $p_T$ hadron-hadron correlations. With an increasing center of mass energies, the nucleus-nucleus collisions happening at LHC focus on characterizing jet-quenching in multi-hadronic final states through a modern jet algorithms \cite{21b, SIBYLL}. 

In this concern, colliders under consideration \cite{23,24}, as an important target for future developments are anticipated to support high luminosity LHC programs. Physical motivations for such  projects are summarized in Ref. \cite{25b,25}. Considerations towards the strengths of very high energy hadron colliders are introduced since long ago in classic pre-phase of SSC-EHLQ \cite{26}.  Recently, an associated framework to study the luminosity of accelerators \cite{27} is further explored towards the understanding of hadrons.

As far a heavy ion collisions and subnuclear matters are concerned, it is worth recalling that nuclear reactions support studying the properties of various nuclear matters and their formation from microscopic constituents. Therefore, our focus lies on the fluctuation structures towards matter formations, and particle productions and their variations under high energy collisions. In this case, one measures the pattern of colliding particles how they fly out from the collisions, for example, the collision of two protons in heavy nuclei. In particle colliders, such collisions include the physics at the Relativistic Heavy Ion Collider (RHIC) as at the Brookhaven National Laboratory \cite{rhic} and the Large Hadron Collider (LHC) at the CERN laboratory in Switzerland \cite{cern}.

We focus on fluctuations that gluons are the glue-like particles, binding the quarks within baryons and mesons, see Mutta {\em et al.}(2009) for an introduction. Quarks make up atomic nuclei, whose stability plays an important role in establishing the properties of matter. Without direct/ indirect measurements, we can not study how the gluons are distributed within an individual proton and dynamical parton distributions \cite{20b}. Following the same, we propose an intrinsic parametric approach to examine the stability or formation of hadronic matter. Experiments at Brookhaven National Laboratory (BNL) \cite{BNL} and the nuclear research at CERN \cite{cern} suggest that the distribution of gluons within a proton fluctuates strongly as a function of time. Hereby, we have investigated the corresponding parametric behavior in our previous study \cite{Ourgluonpaper}. 

Considering that a proton is nearly spherical in its shape, if we take snapshots of the protons in time, each of them would appear radically different. Subsequently, physicists at BNL have considered a model of gluon density function whose fluctuations we have studied in \cite{Ourgluonpaper}. Following the same, in this paper, we focus on string fragmentation and multiple soft interactions. Our analysis supports to interpret new set of data from collision experiments as snapshots of the behavior of protons at a given instant of time \cite{1}. In the above framework, below, we study hadron density fluctuations.
\section{Reggion-Pomeron soft amplitude}

In this section, we provide an extension of the soft interaction model as consider in the framework of the SIBYLL 2.1 \cite{SIBYLL}. Namely, we discuss the stability analysis of soft profile functions under variations of the soft region as well as higher order contributions of the variance of the soft region as a function of the energy and effective particle densities. 

In contrast to the hard interactions that are approximately point-like in their character, we focus on low transverse momentum partons whose interactions are spreaded over a definite region. Such interactions are basic characteristics of soft interactions that involves a nonzero impact parameter $b$. Let the transverse momentum of partons be $p_T$, then the change in $p_T$ and $b$ satisfy the uncertainty principle $\Delta p_T \Delta b \sim 1$. 

In this setting, for a given hadron-hadron collision, see \cite{SIBYLL} for the notations and conventions being used, the energy dependent soft interactions are described by the following soft profile function \\
\begin{eqnarray}
A_{yz} (s,b) = \int d^2b_1 d^2b_2 d^2b_3A_y (b_1)A_z(b_2)A^{soft}(s,b_3) \delta^{(2)}{(b_1-b_2+b_3-b)}
\end{eqnarray}
Here, the mass shell condition is described by $b_1+b_3=b_2+b$, where $b$ is the impact parameter of the interactions. As proposed in SIBYLL $2.1$ \cite{SIBYLL}, soft interactions are described by a fuzzy area, whereby the soft profile function $A^{soft}(s,b_3)$ is arises as an energy dependent Gaussian profile
\begin{eqnarray}
A(s,b_3)=\frac{1}{4\pi B_s(s)}\exp{\bigg[-\frac{|b_3|^2}{4B_s(s)}\bigg]},
\end{eqnarray}
where the variance of the impact parameter concerning the soft interactions is given by \\
\begin{eqnarray}
B_s(s) = B_0+ \alpha^{'} \ln{\bigg(\frac{s}{s_0}\bigg)}
\end{eqnarray}
Here, $s$ is the running mass energy square, and $s_0$ is the initial mass energy square of the constituent hadrons. In this concern, the on-shell amplitude of soft interactions is given by \\
\begin{eqnarray}
A^{soft} (s, b-b_1+b_2)=\int d^2 b_3A^{soft} (s,b_3)\delta^{(2)}(b_1-b_2+b_3-b)
\end{eqnarray}
This amplitude is an overlap between $|in>$ and $|out>$ states of soft interactions in a given band of the energy that reads as
\begin{eqnarray}
Amp=\int dV<\psi_{in}|A|\psi_{out}>
\end{eqnarray}
In a narrow region of energy conservation with aforementioned uncertainty relation, we have the following amplitude
\begin{eqnarray}
Amp =\int dV<A_y(b_1)|A^{soft}_{s,b_3}|A_z(b_2)>
\end{eqnarray}
Using the energy conservation relation and above input and output profile functions, it follows that the underlying amplitude reads as
\begin{eqnarray}
Amp =\int d^2b_1d^2b_2d^2b_3A_y(b_1)A_{z}(b_2) A^{soft}(s,b_3)\delta^{(2)}(b_1-b_2+b_3-b)
\end{eqnarray}
Substituting the above values of input and output profile functions as $A_y(b_1)$ and $A_z(b_2)$ along the $y$ and $z$ directions respectively and the amplitude as the above soft amplitude $A^{soft}(s,b_3)$, the above amplitude is given by 
\begin{eqnarray}
Amp &=& \int d^2{b_1}d^2{b_2} \bigg[  \frac{1}{4\pi {B_p}}exp\bigg(-\frac{b_1^2}{4B_p}\bigg)  \frac{1}{4\pi B_p} \exp\bigg(-\frac{b_2^2}{4B_p}\bigg) \times \nonumber \newline \\ && \frac{1}{4 \pi B(s)_s}\exp-\bigg(\frac{|b-b_1+b_2|^2}{4B_s}\bigg)\bigg]  \nonumber \newline \\
&=& \int \frac{d^2b_1d^2b_2}{(4 \pi )^3 B_p^2 B_s(s)} \exp\bigg[{-\frac{b_1^2}{4B_p}-\frac{b_2^2}{4B_p}-\frac{|b-b_1+b_2|^2}{4 B_s}\bigg]}
\end{eqnarray}
Using the transformations  
\begin{eqnarray}
b_1&=& r_1 \exp {i \theta_1}, \\ \nonumber \newline
b_2&=& r_2 \exp {i \theta_2},
\end{eqnarray} 
the on-shell amplitude
\begin{eqnarray}
A^{soft}(s,b) = \int d^2b_1 d^2b_2A_y (b_1)A_z(b_2)A^{soft} (s,b+b_1+b_2)
\end{eqnarray}
can be expressed as 
\begin{eqnarray}
A^{soft}(s,b) = \int d^2b_1 d^2b_2 \frac{1}{(4 \pi)^3 B_p^2 B_s} \exp{-\bigg[B_s(b_1^2+b_2^2)+B_p | {b-b_1 +b_2|^2}\bigg]}
\end{eqnarray}

With transformations $b_1= r_1 \exp {i \theta_1}$, and $b_2= r_2 \exp {i \theta_2}$, the corresponding measures can be expressed as
\begin{eqnarray}
d^2b_1&=& i r_1 dr_1 \ d \theta_1 \exp^{i\theta_1}, \\ \nonumber \newline
d^2b_2&=& i dr_2 \ d \theta_2 \exp^{i\theta_2}
\end{eqnarray} 
Substituting the above values in the amplitude, the $b_1$ integration is performed by substituting $Ab_1+B = k$ with appropriate changes of its limits. The resulting expression is integrated with respect to $b_2$, wereby by neglecting the higher order term in the $b_1$ integral, we find that the above amplitude can be expressed as
\begin{eqnarray}
A^{soft}(s,b) &=&
\frac{1}{A^2}\bigg[{exp(-B^2)} -B\bigg(1-\frac{exp(-k^2)}{\sqrt \pi k}(1-\frac{1}{2k^2})\bigg)\bigg] \times \nonumber \newline\\&& 
\frac{4 \pi^2}{A^2} \bigg[\int b_2\exp\bigg(-\frac{b_2^2}{4B_p}\bigg) \exp\bigg(-\frac{b^2+b_2^2+2bb_2}{4B_s}\bigg)  \times \nonumber \newline\\&&
exp\bigg( \frac{b^2+b_2^2+2bb_2}{16 B_s^2}\bigg)db_2\bigg]
\end{eqnarray}	
where the model parameters $A$ and $B$ are given by
\begin{eqnarray}
A &=& \sqrt{\frac{1}{4B_p}+ \frac{1}{4B_s}}, \nonumber \newline \\
B &=& \frac{b+b_2}{4B_s}\frac{1}{\sqrt{{\frac{1}{4B_p}+ \frac{1}{4B_s}}}}.
\end{eqnarray}

In the limit of small $B_s(s) \rightarrow 0$, the soft profile function becomes a delta function, whereby the above amplitude results into hard profile function. Using the above approximation of $B_s$ and first order estimated value of the input and output profile functions. In this case, one uses data \cite{SIBYLL} to fit the parameter $B_p$ that characterizes the transverse size of proton gives an energy dependent Gaussian profile function in terms of the impact parameter $b$. In other words, for a given proton-proton collision, it follows that the above integral can be represented \cite{SIBYLL} by the following Gaussian profile
\begin{eqnarray}
A_{pp}^{soft}(s,b) = \frac{1}{4 \pi(2B_p+B_s)}\exp\bigg[-\frac{b^2}{4(2B_p+B_s)}\bigg]
\end{eqnarray}
Physically, this consideration is analogous to a fuzzy area that is in late time analysis is parameterized by a Gaussian energy dependent soft profile. In practice, the expansion of the underlying variance involve the Regge trajectory whereby its Taylor expansion coefficient are determine by derivatives of the Regge trajectory. The first derivative terms allows to include motion of nucleons, its second derivative described accelerated nucleon and the third derivative describes non uni-formally accelerated nucleons with jerk, and other higher order terms as discussed in section $6$ below. The analysis of higher order correction in Regge trajectory is clearly visible when the expansion is performed in the logarithmic scale of the running mass energy square with respect to a given initial energy scale.
\section{Multiple soft hadronic fluctuations}

In this section, we study fluctuation properties of hadron-hadron collisions in relation to Lund string fragmentation, multiple soft interactions and their Gaussian limits. Below, we provide the fluctuation theory based stability examinations of the soft models.
\subsection{Lund string fragmentation}
In this subsection, we consider an intrinsic analysis of the model qualitatively under fluctuations of the model parameters. Given the quark and diquark energy density function $A(\alpha, \mu)$, its fluctuations are described through the embedding function reading as 
\begin{equation} \label{fql}
A:\Sigma \rightarrow \mathbb{R},
\end{equation}
Namely, in the light of the fluctuation theory, it is given by the quark or diquark energy fragmentation as the following distribution
\begin{eqnarray} \label{qdo}
A= \frac{(1-x)^\alpha}{(x^2+\frac{\mu^2}{s})^\frac{1}{4}}
\end{eqnarray}
Under variations of the model parameters $\{\alpha, \mu\}$, we find that the quark/ diquark distribution $A(\alpha, \mu)$ has the following flow components
\begin{eqnarray}
\frac{\partial A}{\partial \alpha} &=& \frac{(1-x)^\alpha \log(1-x)}{(x^2 +\frac{\mu^2}{s})^\frac{1}{4}}, \nonumber \newline \\
\frac{\partial A}{\partial \mu} &=& \frac{-(1-x)^\alpha \mu}{2s(x^2+\frac{\mu^2}{s})^\frac{5}{4}}
\end{eqnarray}
Hereby, the flow equation $\frac{\partial A}{\partial \alpha} = 0$ implies that $\alpha= \infty$, whenever $x<1$. However, here $A=0$ itself implies that it is infinity in $\alpha$ direction. Similarly, the equality $\frac{\partial A}{\partial \mu} = 0$ implies that $\mu$=0 as the solution.

The stability analysis is carried around the above critical points that arises as the roots of the flow equations $ A_{\alpha}=0$ and $A_{ \mu} =0$, whereby by jointly evaluating the signs of one of the heat capacities $\{A_{\alpha \alpha}, A_{ \mu  \mu} \}$ and the fluctuation determinant $\Delta: = |H|$ as the first and second principal minors as in the below fluctuation matrix 
\begin{equation} \label{hm}
H:= \frac{\partial^2}{\partial x^i \partial x^j}A(\alpha, \mu)
\end{equation}
for a chosen parameter vector $\vec{x}:= (\alpha, \mu)$ on the fluctuation surface $\Sigma $ of the model parameters $\{ \alpha, \mu\}$.
In this concern, the stability of an ensemble of quarks or diquarks is govern by the second derivatives
\begin{eqnarray}
\frac{\partial^2A}{\partial \alpha^2} &=& \frac{(1-x)^\alpha (\log(1-x))^2}{(x^2+\frac{\mu^2}{s})^\frac{1}{4}}, \nonumber \newline \\
\frac{\partial^2A}{\partial \mu^2} &=& \frac{(1-x)^\alpha (-2sx^2 +3\mu^2}{4s^2 (x^2+\frac{\mu^2}{s})^\frac{9}{4}}, \nonumber \newline \\
\frac{\partial^2A}{\partial x \partial \mu} &=& -\frac{(1-x)^\alpha \mu\log(1-x)}{2s(x^2+\frac{\mu^2}{s})^\frac{5}{4}}
\end{eqnarray}
In relation to the statistical systems, we may hereby classify the nature of QCD-improved soft gluonic configurations with their constituent probability of quarks as in Eqn.(\ref{qdo}) having values of the parameters as
\begin{equation}
\Sigma := \{ (\alpha^{(1)}, \mu^{(1)}), (\alpha^{(2)}, \mu^{(2)}), (\alpha^{(3)}, \mu^{(3)}), \ldots, (\alpha^{(N)}, \mu^{(N)}) \},
\end{equation}
where it corresponds to a thermodynamical basis when we take $N \rightarrow \infty$. The associated global stability of the ensemble under fluctuations of $\{\alpha, \mu \}$, we are required to compute the fluctuation discriminant $\Delta:= A_{\alpha \alpha} A_{\mu \mu} - (A_{\alpha \mu})^2$. Substituting the above expressions for fluctuation capacities 
$\{ A_{\alpha \alpha}, A_{\mu\mu}\}$ and the underlying cross correlation $A_{\alpha \mu}$, we obtain the below fluctuation discriminant
\begin{eqnarray}
\Delta = -\frac{(1-x)^{2\alpha} (sx^2-\mu^2)(\log(1-x))^2}{2s^2 (x^2+\frac{\mu^2}{s})^\frac{5}{2}}
\end{eqnarray}
Therefore, it follows that the only critical point of $A$ is $(-\infty, 0)$. At $b$=0 it follows that we have the following limiting determinant
\begin{eqnarray}\Delta= -\frac{1}{sx^3} \log(1-x).(1-x)^{2\alpha} \end{eqnarray}
On the other hand, for $s<0$, the sign of delta varies according as $\log(1-x)$. Thus, when $x>0$, it follows that we have $\Delta<0$. Similarly, for $x<0$, we have  $\Delta>0$. For $x=1$, it is not difficult to see that $\Delta=0$. Therefore, at $\mu=0$, the fragmentation of strings attaching quark-antiquark and diquark-antidiquark with fractional energy $x$ is stable according as $|\mu|<x\sqrt{s}$, where $s$ is the center of mass energy square. In this region, there is a possibility of production of new particles, i.e., this quark-diquark model has phase transitions, namely, the constituents particles will start decaying at the above or a higher energy scale. 

The corresponding diagram of stability regions is shown in fig. (\ref{lundsf}) below. On the other hand, when there is no string fragmentation, that is, at $x=0$, we have $\Delta=-\infty$. Hence, the system becomes ill-defined. Similarly, at $x=1$, we have an undefined $\Delta$ that is there is no well define distribution of quarks and diquarks. In order to illustrate the above analysis qualitatively, we may choose $x= 1/2$ and $s=1$, whereby we have the following limiting determinant 
\begin{eqnarray}\Delta(\alpha,\mu)= (\log2)^2\frac{\mu^2-1/4}{2^{2\alpha+1}(\mu^2+1/4)} \end{eqnarray}
The  concerned qualitative behavior is shown in the fig. (\ref{lundsf}) below.
In the UV limit, it follows that we have a small values of $\mu$, wherefore in this region, the above determinant$\Delta$ simplifies as  
\begin{eqnarray} \Delta = -\frac{(\log(1-x))^2 (1-x)^{2\alpha}}{sx^3} \end{eqnarray}
In this case, we see that the determinant scale stays as the inverse of the energy scale. Herewith, in the IR limit where the quark effective mass $\mu$ is much larger than the other terms such as the energy scale, we find that the determinant reads as
\begin{eqnarray} \Delta= \frac{\sqrt{s}(\log(1-x))^2}{\mu^3} \end{eqnarray}
In this limit, we see that determinant varies as the square root of the energy scale $s$. Further, the limiting determinant is inversely proportional to the cube of the center of mass energy $\mu$. On the other hand we see in the UV limit that the corresponding determinant becomes the independent of the center of mass energy square $s$. In this case, the system stability is solely dependent on the fact that energy is absorbed. Moreover, in the UV limit, we notice that all the systems emitting energy are observed to be unstable. Note that the above picture is true only in the DPM limit of partons and the formation of quark matter through the process of Lund string fragmentation. 

Similarly, in the IR limit, it follows that new particles are produced when the screening effects are too strong and the quark has an effective negative mass. This process is possible through the vacuum polarization and creation of virtual particles \cite{SCHWINGERb}. In this case, we have found that Lund string fragmentation arises with a class of effective quark masses that are well bounded by the running centre of mass energy. In particular, there is a band of instability for $|\mu|\le \frac{\sqrt{s}}{2}$ as depicted in fig. (\ref{lundsf}). For all $|\mu|> \frac{\sqrt{s}}{2}$, it follows that we have a stable ensemble of fractional momentum quarks and diquarks under variation of the effective quark mass energy $\mu$, see fig.(\ref{lundsf}) for a diagrammatic view. 
\begin{center}
	\begin{figure}
		\hspace*{0.0cm} \vspace*{-1.4cm}
		\includegraphics[width=12.0cm,angle=0]{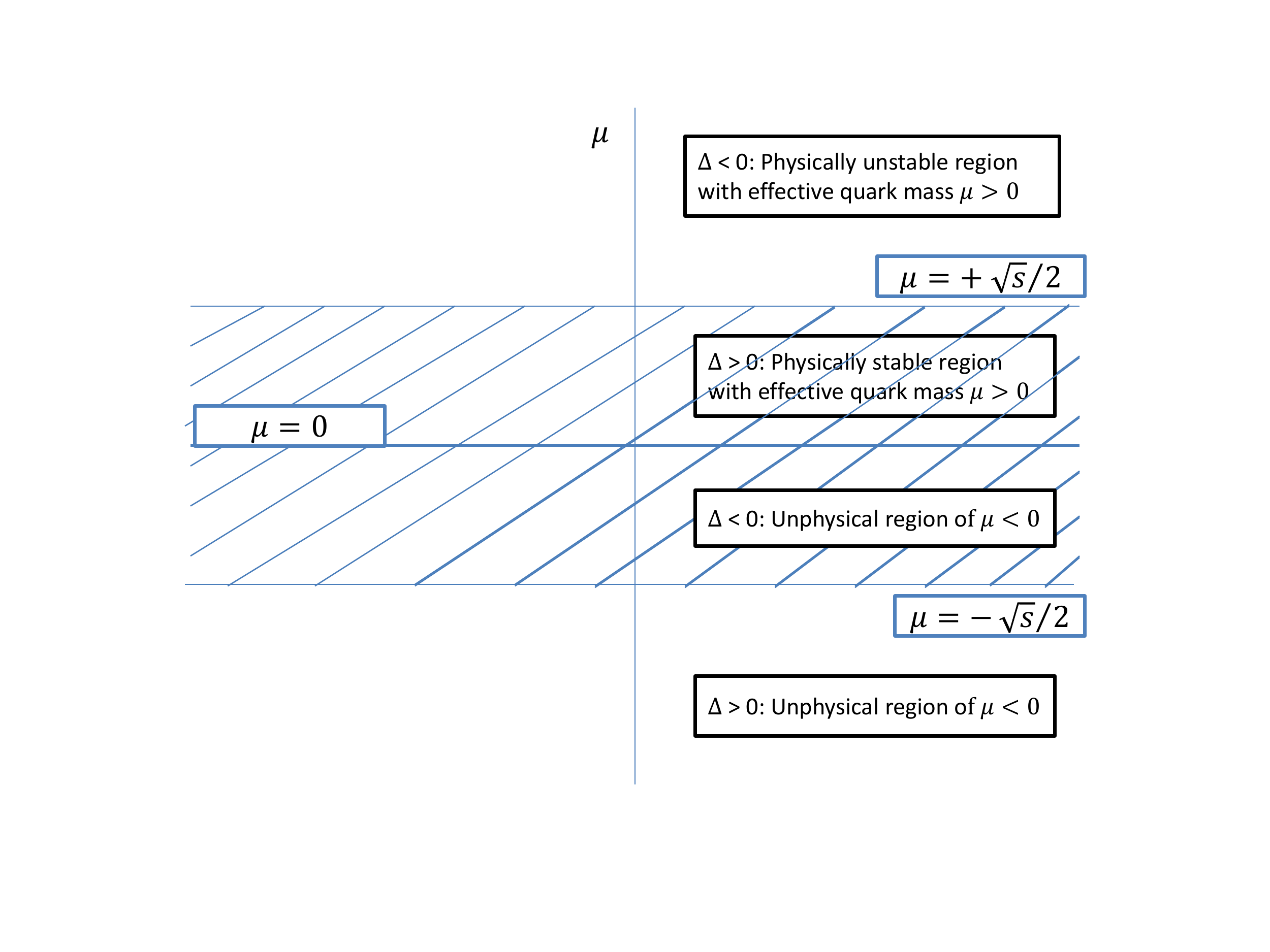}
		\caption{The region of stability plotted as a function of the effective quark mass energy $\mu$ on the $Y$-axis and $\Delta$ on $X$-axis describing the nature of an ensemble of quark-antiquark and diquark-antidiquark pairs by considering variations of the energies $E$ for the values of the model parameters $\alpha= 3.0$ and $\mu=0.35$ GeV. Here, $\Delta<0$ characterizes an unstable region where new particles are produced or decay, $\Delta>0$ characterizes a stable region of quark-antiquark and diquark-antidiquark pairs and the line $\Delta=0$ describes the state where the quark or diquark distribution is too flat.} \label{lundsf}
		\vspace*{-0.01cm}
	\end{figure}
\end{center}
Herewith, it follows that new particles are produced when a quark or diquark distribution function leads to an unstable ensemble. In this case fluctuations continue to persist until the mass difference between the produced and incoming states for smaller than the string threshold mass. In particular, when the fluctuations are stopped final hadrons are formed in the characteristics of our model.  Hereby, we have determine the optimal region of the effective quark mass where new particles are expected to be produced. Namely, under fluctuations of the model parameters, our analysis is sensitive to determine the nature of primodial transverse momentum of parent quarks or diquarks, i.e., a function of the center of mass energy of hadron-hadron interactions. Note that in the setup of Lund string fragmentation model, if the quark has a stable probability distribution, then the diquark will have an unstable probability distribution in the limit of DPM picture. This finds high importance in QCD improved parton models as well as in the associated strong coupling fluctuations concerning the proton-proton or proton-antiproton interactions.
\subsection{Lund string fragmentation in the Gaussian limit}
In the case of Gaussian limit, the Lund string fragmentation model with a primordial equal magnitude opposite sign transverse momentum pair satisfies a Gaussian distribution profile for a given center of mass energy square of hadron-hadron interactions. The corresponding energy fractionation of each new particle produced satisfies a Gaussian profile as a function of the transverse mass. Here, the fractional energy of each new particle with respect to parent quarks or diquarks depends on the variance of the distribution. 

In particular, the fragmentation starts when the threshold mass becomes comparable to the resulting string mass. In this case, the corresponding string finishes the fragmentation of quarks or diquarks by forming to hadrons as the final state. It is found that this sample has a positive auto-correlation as well as a positive cross-correlation for all $z$ in $(0,1)$, where $z$ is fraction of energy of new particle with respect to original quarks and diquarks. For a primordial pair of quarks or diquarks of opposite sign and equal transverse momentum mediated in the above dual parton picture of the limiting minijet model, as a function of the model parameters $a$ and $b$, the corresponding Gaussian probability density of quarks or diquarks is given by 
\begin{eqnarray} \label{qd}
f(a, b)= \frac{(1-z)^a }{z}\exp\bigg[-\frac{bm_T^2}{z}\bigg]
\end{eqnarray}
To examine the local fluctuations of the above limiting Gaussian density $f(a, b)$, we compute the corresponding flow components of $f(a, b)$. In this case, we observe that the corresponding $a$ and $b$-flow components can respectively be written as
\begin{eqnarray}
\frac{\partial f}{\partial a} &=& \frac{(1-z)^a\log(1-z)}{z}\exp\bigg[- \frac{bm_T^2}{z}\bigg], \nonumber \newline \\
\frac{\partial f}{\partial b} &=& \frac{m_T^2(1-z)^a}{z^2}\exp\bigg[- \frac{bm_T^2}{z}\bigg]
\end{eqnarray}
In order to investigate the nature of stability of the constituent quark probability density profile as in Eqn.(\ref{qd}), we need to compute the fluctuation capacities defined as its second pure derivatives with respect to the model parameters $\{ a, b \}$. 
In this direction, we notice that the pure capacities of the above density profile $f$ simplify as
\begin{eqnarray}
\frac{\partial^2f}{\partial a^2}&=&\frac{(1-z)^a\log(1-z)^2}{z}\exp\bigg[- \frac{bm_T^2}{z}\bigg], \nonumber \newline \\
\frac{\partial ^2f}{\partial b^2}&=&\frac{m_T^4(1-z)^a}{z^3}\exp\bigg[- \frac{bm_T^2}{z}\bigg]
\end{eqnarray}
Furthermore, it follows that the corresponding cross-correlation capacity reads as\begin{eqnarray}\frac{\partial^2f}{\partial a \partial b} &=& -\frac{m_T^2(1-z)^a\log(1-z)}{z^2}\exp\bigg[- \frac{bm_T^2}{z}\bigg] \end{eqnarray}
It is worth mentioning that the system corresponds to an indefinite statistical configuration because the underlying determinant $\Delta$ vanishes identically under fluctuations of the parameters $\{ a, b\}$, viz. we have $\Delta$ = 0. In the above setup, we observe that there may or may not be phase transitions in this limit that is because of the fact that a high degree of symmetry in the probability distribution of quarks and diquarks that arise as a long term effect. Notice that if the quark probability distribution is maximized, then the corresponding diquark probability distribution gets minimized because of the same the probability distribution satisfy \cite{SIBYLL} the flowing conservation law 
\begin{eqnarray}
f_q+f_{qq}=1
\end{eqnarray}
\subsection{Multiple soft interactions}
Under fluctuations of the Reggeon, Pomeron and scale of initial mass energy square, we study stability analysis of a system with multiple soft interactions. Here, we consider the scattering cross section $A$ as computed in the minijet model \cite{SIBYLL}, whereby we consider it as an embedding from the space of Reggeon-Pomeron parton densities and the initial energy scale to the set of real numbers as
\begin{eqnarray} \label{rpsc} A = X \bigg(\frac{s}{E}\bigg)^p + Y\bigg(\frac{s}{E}\bigg)^{-r} \end{eqnarray}
Notice that the Reggeon exchange as obtained at low energy is found by numerical fits \cite{31b}. However, the Pomeron exchange density depends on decomposition of Pomeron terms into hard and soft contributions. Physically, this shows that the Pomeron density arises as a function of transverse momentum. In the QCD improved parton model, the transvere momentum is a non-trivially energy dependent function which in the minijet limit simplifies as 
\begin{eqnarray}
p_T^{min}  = p_T^0 + \lambda \exp{c(\sqrt{\ln(s/GeV^2)})}, 
\end{eqnarray}
where initial parameters are taken as $p_T^0 = 1 GeV, \lambda = 0.065 GeV, c= 0.9$ and $\sqrt{s_0}$ is taken in a varied range of energies from $1$ GeV to $1800$ GeV, involving various phenomenological models and experiments as discussed in the next section. 

To study fluctuations of the above multiple soft interaction density, we need to compute the corresponding flow components of the energy dependent Reggeon-Pomeron scattering cross section as the model embedding function
\begin{equation} \label{fql}
A:\mathcal{M}_3 \rightarrow \mathbb{R},
\end{equation}
where the manifold $\mathcal{M}_3$ is locally coordinatized by a triple $(E, p, r)$ through the function $A$ as depicted in the above Eqn.(\ref{rpsc}). In this case, concerning the multiple soft interactions, it follows that the local flow components can be written as
\begin{eqnarray}
\frac{\partial A}{\partial E} &=& - X\frac{p}{E} \bigg(\frac{s}{E}\bigg)^p + Y \frac{r}{E} \bigg(\frac{s}{E}\bigg)^{-r}
, \nonumber \newline \\
\frac{\partial A}{\partial p} &=& X \bigg(\frac{s}{E}\bigg)^p \log\bigg(\frac{s}{E}\bigg), \nonumber \newline \\
\frac{\partial A}{\partial r} &=& - Y\bigg(\frac{s}{E}\bigg)^{-r} \log\bigg(\frac{s}{E}\bigg)
\end{eqnarray}
Herewith, the parametric fluctuation stability of the ensemble is described by the first, second and third principal minors of the below Hessian matrix 
\begin{equation} \label{hm}
H:= \frac{\partial^2}{\partial x^i \partial x^j}A(E, p, r)
\end{equation}
for the chosen parameter vector $\vec{x}:= (E, p, r)$ on the surface $\Sigma_3 $ of parameters $\{E, p, r \}$. 
In the light of statistical mechanics, we wish to classify the nature of an ensemble of QCD-improved Reggeon-Pomeron configurations with their scattering cross-section as in Eqn.(\ref{rpsc}) having values of the parameters as
\begin{equation}
\Sigma_3 := \{ (E^{(1)}, p^{(1)}, r^{(1)}), (E^{(2)}, p^{(2)}, r^{(2)}),, \ldots, (E^{(N)}, p^{(N)}, r^{(N) }) \}
\end{equation}
Here, the model corresponds to a (non)interacting thermodynamical system according as the values of the correlation, where we take $N \rightarrow \infty$. The fluctuations capacities concerning the above model embedding as depicted in Eqn.(\ref{rpsc}) are given by its pure second derivatives
\begin{eqnarray}
\frac{\partial^2A}{\partial r^2} &=& Y \bigg(\frac{s}{E}\bigg)^{-r} \bigg(\log(\frac{s}{E})\bigg)^2, \nonumber \newline \\
\frac{\partial^2A}{\partial p^2} &=& X \bigg(\frac{s}{E}\bigg)^p \bigg(\log\frac{s}{E}\bigg)^2, \nonumber \newline \\
\frac{\partial^2A}{\partial E^2} &=& X s^p p(p+1)E^{-p-2}+Yr(r-1) s^{-r} E^{r-2}
\end{eqnarray}
The associated local correlations concerning the above embedding as in Eqn.(\ref{rpsc}) are given by it's mix derivatives as per the following data
\begin{eqnarray}
\frac{\partial^2A}{\partial E \partial p} &=& -X s^p E^{-p-1}\bigg(p \log(\frac{s}{E})+1\bigg), \nonumber \newline \\
\frac{\partial^2A}{\partial E \partial r} &=& Y E^{r-1} s^{-r}\bigg(r \log(\frac{E}{s})+1\bigg), \nonumber \newline \\ 
\frac{\partial^2A}{\partial p \partial r} &=& 0
\end{eqnarray}
Under the variation of Reggeon and Pomeron densities we see that Pomeron capacity $A_{pp}$ remains positive when $p$ is even number. In the case, when $p$ is an odd number $A_{pp}$ is positive whenever $s$ and $s_0$ have the same size. In this direction, we observe that Reggeon capacity $A_{rr}$ obeys the same convention of stability as that of $A_{pp}$ with replacement of $p$ by $r$. On the other hand, we see that the energy fluctuation capacity $A_{EE}$ remains positive, whenever we have 
\begin{eqnarray}
\frac{X}{Y}(\frac{s}{E})^{p+r} > \frac{r(1-r)}{p(1+p)}
\end{eqnarray}	
Similarly, it follows that correlation of the Pomeron with energy vanishes if the running mass-energy square satisfies an exponential decaying relation in $1/p$, i.e., we have the following scaling
\begin{eqnarray} s(p) = E e^{-1/p} \end{eqnarray}
On the other hand, there is no correlation of Reggeons with initial energy square if it's running mass-energy square $s$ satisfies \begin{eqnarray} s(r) = E e^{1/r} \end{eqnarray} Hereby, we observe that Pomeron and Regeoon have opposite correlation property, namely, if one get an increasing correlation with energy, the other gets a decreasing correlation with energy square. As far as the minijet model with multiple soft interactions are concerned, we see that the Reggeons and Pomerons are uncorrelated, that is, we have $A_{pr}=0$. 

At this juncture, under simultaneous fluctuations of the energy square, and Reggeon and Pomeron densities stability is determined by computing the determinant and related principal minors of the Hessian matrix 
\begin{eqnarray} 
M= \left[ {\begin{array}{ccc}
	A_{EE}& A_{Ep}&A_{Er} \\
	A_{Ep}&A_{pp} &A_{pr} \\
	A_{Er}&A_{pr} &A_{rr} \\
	\end{array} } \right]
\end{eqnarray}   
In other words, by applying elementary operations $C_1 \rightarrow C_1-C_3\frac{A_{Er}}{A_{rr}}$ and then $C_1 \rightarrow C_1-C_2\frac{A_{Ep}}{A_{pp}}$, with $p_3$ as given below in Eqn.(\ref{minors3}), we see that the above Hessian matrix reduce as the triangular matrix 
\begin{eqnarray} 
M= \left[ {\begin{array}{ccc}
	p_{3}& A_{Ep}&A_{Er} \\
	0 &p_{2} & 0 \\
	0 & 0 &p_{1} \\
	\end{array} } \right] 
\end{eqnarray} 
Hereby, we find that the above principal minors $p_1$ $p_2$ and $p_3$ are given by 
\begin{eqnarray} \label{minors3}
p_1& = &A_{rr}, \nonumber \newline \\
p_2& = & A_{pp}, \nonumber \newline \\
p_3&= & A_{EE}-\frac{A_{Ep}A_{Er}}{A_{rr}} -\frac{ (A_{Ep})^2}{A_{pp}}
\end{eqnarray} 
In this concern, it is worth mentioning that the system of Reggeon and Pomeron remains stable as long as we have positive values of $p_1$, $p_2$ and $p_3$. It is clear that the sign of $p_1$ and $p_2$ remains the same as the sign of $A_{rr}$ and $A_{pp}$. Further, $p_3$ is positive whenever we have the following inequality
\begin{eqnarray} 
\frac{A_{EE}}{A_{Ep}}> \frac{A_{Er}}{A_{rr}} + \frac{A_{Ep}}{A_{pp}}
\end{eqnarray} 
In this framework, the critical points of the model flow are given by 
\begin{eqnarray} 
\frac{\partial A}{\partial E}=0= \frac{\partial A}{\partial p} = \frac{\partial A}{\partial r} \end{eqnarray}
It is not difficult to see that the solution to above equations are given by  \begin{eqnarray} \frac{X}{Y}(\frac{s}{E})^{p+r} = \frac{r}{p} \end{eqnarray} as the solution to $\frac{\partial A}{\partial E}=0$ and  $s=E$ as a repeated root of the other two flow equations involving Reggeons and Pomerons respectively. Thus, the simultaneous solution to above flow equations is given by \begin{eqnarray} \frac{X}{Y} = \frac{r}{p}.\end{eqnarray} In other word, we see that the critical theory is independent of the initial energy square $s_0$. For $X=Y=1$ and $s=\lambda E$, the minor $p_3$ simplifies as \begin{eqnarray} 
p_3=\frac{1}{E^2}\lambda^p[(p(p+1)-\frac{r+p}{\ln\lambda}(p\ln\lambda+1)+\lambda^{-r-p }r(r-1)]\end{eqnarray}
In the special case of $r=1$, it follows that we have $p=-1$ whenever $p_3=0$. However, in the converse case of $p=-1$, we notice that the equation $p_3=0$ simplifies as a nonlinear equation in $r$ as \begin{eqnarray} \label{nle1} 1-\ln \lambda+r\lambda^{-1-r} \ln \lambda = 0 \end{eqnarray}
Here, the case of $r=1$ is indeed a solution for $p_3=0$ with $p=-1$. However, it is possible to find other values of $r$ that satisfy the above Eqn.(\ref{nle1}). An approximate solution of $r$ can be found by solving this equation numerically for a given energy scale $\lambda$. 
For the case of $r=0$, we see that the minor $p_3$ vanishes according as the vanishing of $p\ln\lambda-p$. This is possible if, for $p\ne 0$, we have $\lambda$ = $e$. Thus, for $\lambda<e$, it follows that we have $p_3>0$. Similarly, and for $\lambda>e$, it is not difficult to see that we have $p_3<0$. 
\begin{center} 
	\begin{figure}
		\hspace*{-1.0cm}	\vspace*{-8.0cm} 
		\includegraphics[width=15.0cm,angle=0]{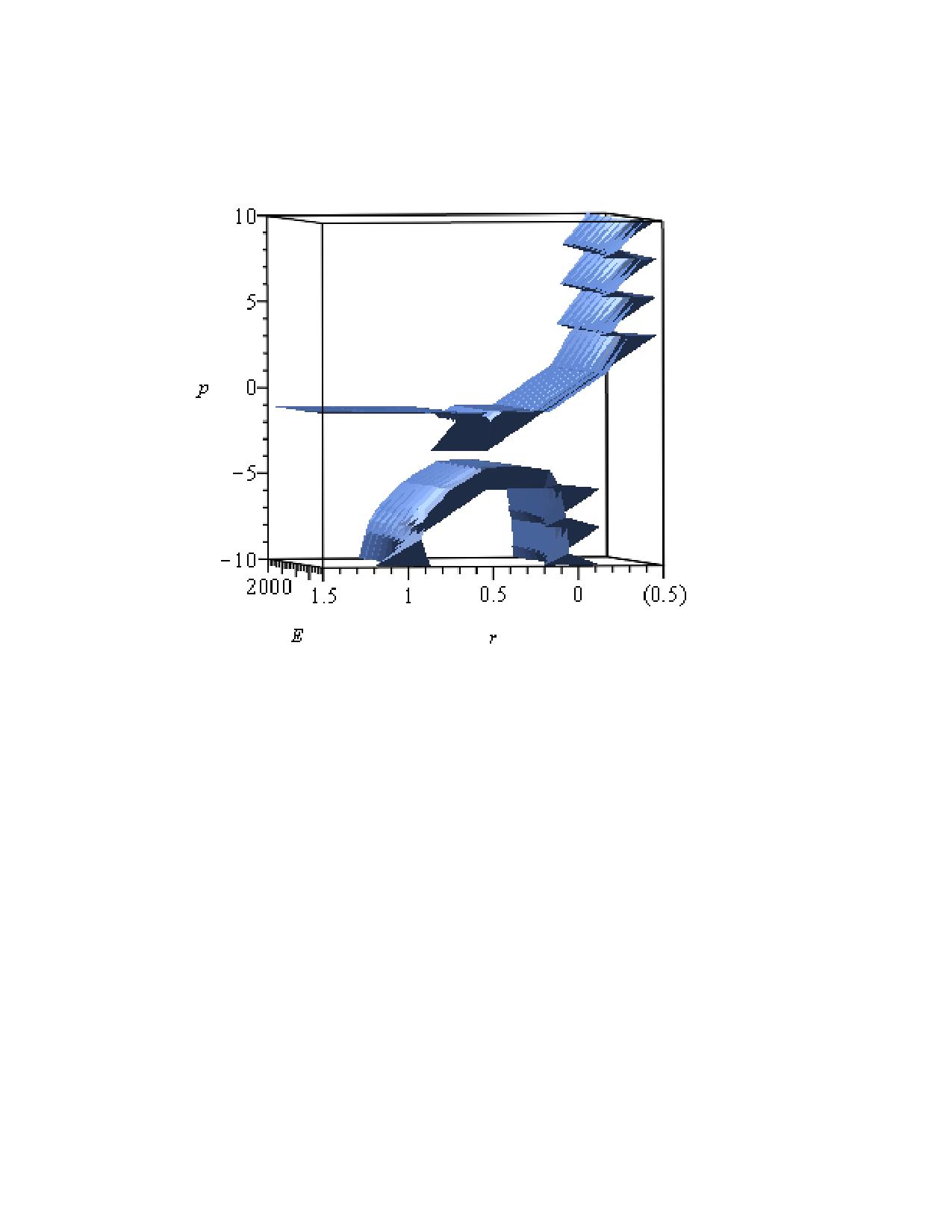}
		\caption{The region of stability under fluctuations of the Reggeon and Pomeron densities surface-plotted as a function of the Reggeon and Pomeron densities $\{r, p\}$ on the $X$-axis and $Y$-axis respectively and the energy scale on $Z$-axis describing the nature of the ensemble of Reggeons and Pomerons by considering variations of the energies $E$ for its values  $1-2000$ GeV. Here, $\Delta<0$ characterizes unstable region where new particles are prone to be produced, and $\Delta>0$ characterizes the stable region of Reggeons and Pomerons, and the line $\Delta=0$ describes the state where the soft scattering cross section become too flat in $\{E, p, r\}$ space.} \label{p3fig}
		\vspace*{7.01cm}
	\end{figure}
\end{center} 	
\vspace*{-1.01cm}
\begin{center} 
	\begin{figure}
		\hspace*{-1.0cm}	\vspace*{-8.0cm} 
		\includegraphics[width=15.0cm,angle=0]{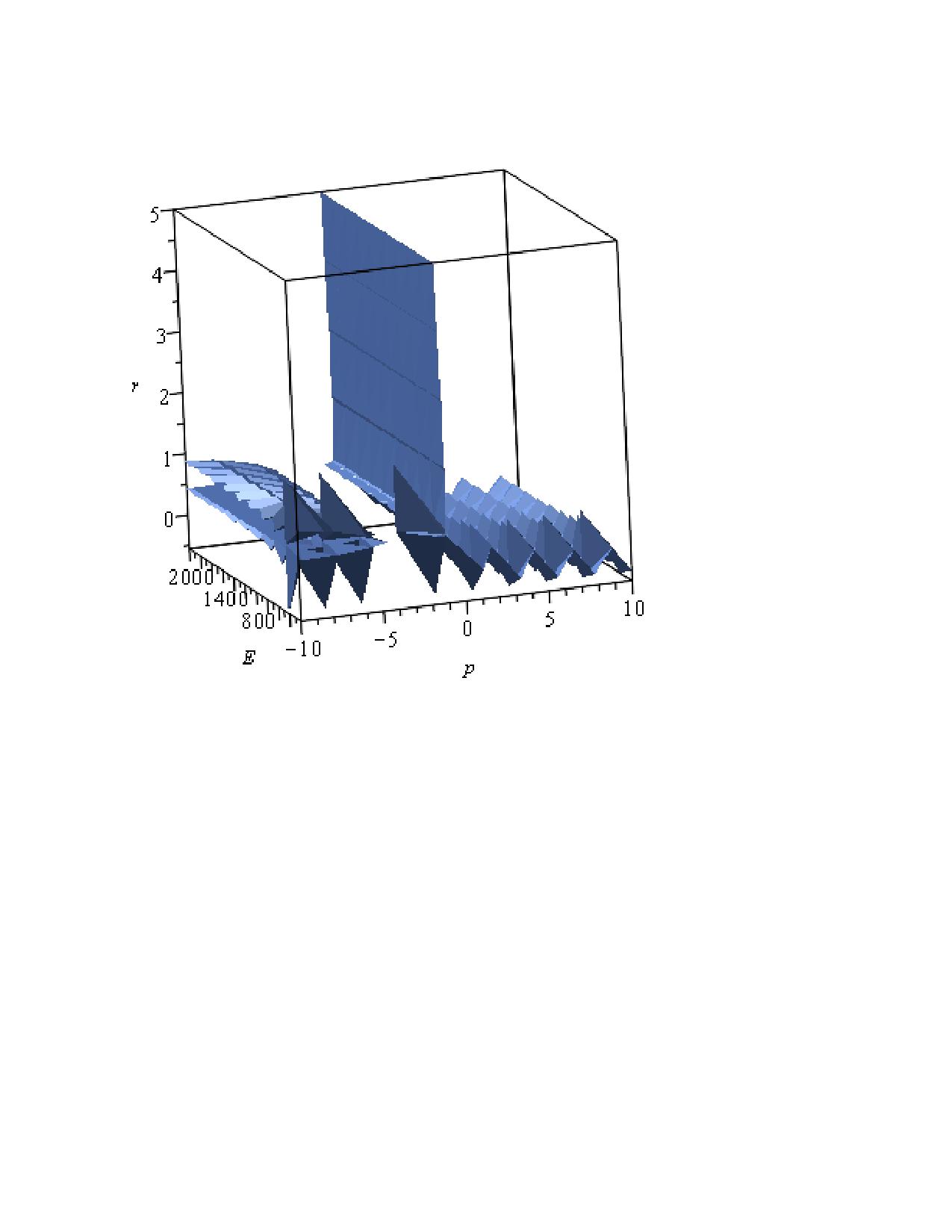}
		\caption{The Pomeronic view of region of stability as the surface-plot as a function of the Reggeon and Pomeron densities $\{r, p\}$ on the $X$-axis and $Y$-axis respectively and energy scale on $Z$-axis describing the nature of the ensemble of Reggeons and Pomerons by considering variations of the energies $E$ for its values $1-2000$ GeV. Here, $\Delta<0$ characterizes unstable region where new particles are prone to be produced, and $\Delta>0$ characterizes the stable region of Reggeons and Pomerons and the line $\Delta=0$ describes the state where the soft scattering cross section become too flat in $\{E, p, r\}$ space.} \label{p3}
		\vspace*{7.01cm}
	\end{figure}
\end{center} 	
\vspace*{-1.01cm}
Further, for the choice of $p=0$ with $\lambda=\frac{s}{E}$, it follows that, for $p_3=0$, we have the below the below nonlinear relation \begin{eqnarray} -r+\lambda^{-r} \ln\lambda r (r-1) = 0. \end{eqnarray} In other words, the curve separating the regions of stability and instability, that is termed as the wall of stability/instability is given by 
\begin{eqnarray}
r=1+\frac{\lambda^{-r}}{\log\lambda}.\end{eqnarray} Finally, for $p=-r$, the separation boundaries are given by $p(p+1)=0$. Hence, we see that the value of Pomeron-Reggeon densities are $(0,0)$ and $(1,-1)$ respectively as the possible values for $(r,p)$. In general, for $r=k, p=\frac{Y}{X}k$ and $E=s$, we find that the minors of the Hessian matrix as depicted in Eqn.(\ref{hm}) are given by $\frac{Yk^2}{Xs^2}(X+Y)$, $-\frac{X^2}{s^2}$ and $0$ respectively.

In fig.(\ref{p3fig}), we provide surface plot of $p_3=0$ with $X=Y=1$ to illustrate the qualitative nature of parametric fluctuations. Similarly, for the linear minor $p_1>0$, we must have same sign for $s$ and $E$ for $Y>0$ with $r$ an odd number. On the other hand, if $Y>0$ and the sign of $s$ and $E$ is opposite, then all the principal minors $\{ p_1,p_2,p_3 \}$ become complex numbers. As far as the sign is concerned, the same conclusions holds for the surface minor $p_2$ that of the linear minor $p_1$ with a mere replacement of $Y \rightarrow X$ and $r \rightarrow p$. Hereby, it is not difficult to verify that the determinant of the  Hessian matrix defined as the product $\Delta=p_1p_2p_3$ simplifies as the below expression
\begin{eqnarray}	
\Delta&=& XY  (\frac{s}{E})^{p-r}(\ln\frac{s}{E})^4 \bigg[\frac{X}{E^2}(\frac{s}{E})^p \{p(p+1)-\frac{r+p}{\ln\frac{s}{E}} (p\ln\frac{s}{E}+1)\} \nonumber \\ \newline & +& \frac{Yr(r-1)}{E^2} (\frac{s}{E})^{-r} \bigg]
\end{eqnarray}

\begin{center} 
	\begin{figure}
		\hspace*{-1.0cm}
		\includegraphics[width=15.0cm,angle=0]{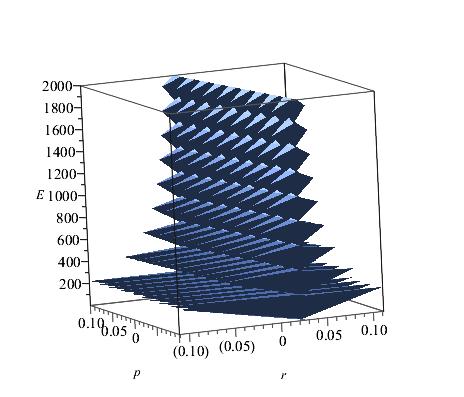}
		\caption{The energy view of region of stability as the surface-plot, depicted as a function of the Reggeon and Pomeron densities $\{r, p\}$ on the $X$-axis and $Y$-axis respectively, and the energy scale on $Z$-axis describing the nature of an ensemble of Reggeons and Pomerons by considering variations over the energies $E$ for its values in $1-2000$ GeV. Here, the determinant $\Delta<0$ characterizes an unstable region where new particles are prone to be produced, and $\Delta>0$ characterizes the stable region of Reggeons and Pomerons and the line $\Delta=0$ describes the critical state where the soft scattering cross section become too flat in the space of $\{E, p, r\}$.} \label{p3energyview}
		\vspace*{7.01cm}
	\end{figure}
\end{center} 	
\vspace*{-.5cm}
The particular behavior of $ \Delta$ whether a given hadronic system is stable or prone to produce new particles is discussed as below. In this case, we notice that the determinant $\Delta$ arises as the volume minor.
In the sequel, following the analysis of marginal wall of instability as introduced by Sen \cite{As} in the case of black holes, we discuss the wall of stability/instability for the case of $r=p$ as the special case of $\Delta=0$. In this case, we find that the concerned marginal wall is described by $p_3$ (or similarly that of $\Delta$ up to a factor) is given by  
\begin{equation}
Xs^{r}E^{-r-2}[r(r+1)-\frac{2r}{\ln \frac{s}{E}}(r\ln \frac{s}{E}+1)]+Y[r(r-1)s^{-r} E^{r-2}]
\end{equation}
Our analysis is performed on the fundamentals of minijet model, where the scattering cross section is described as a function of the initial mass energy square, and Reggeon and Pomeron densities in the concern of proton-proton and proton-antiproton interactions. In general, at the scale of  $s=\lambda E$, we see that the wall of stability/ instability arises as the curve  
\begin{equation} \label{rnep}
p_3=X\lambda^r E^{-2}[r(r+1) -\frac{2r}{\ln\lambda}(r\ln\lambda+1)]+Yr(r-1)\lambda^{-r }E^{-2}			\end{equation}
This is the limiting wall of (in)stability for the case of $r=p$. This solely depends on $r$ and $E$ for given coefficients $X$ and $Y$ as the Pomeron and Reggeon coefficients respectively. In the special case of $\lambda=e$ and $X=1=Y$, from the above Eqn.(\ref{rnep}), it follows that we have
\begin{equation} \label{rps}
p_3= -\frac{2r}{E^2}\bigg(r\ sinh(r)+ cosh(r)\bigg)
\end{equation}
In this particular case, for a positive $r$, we see that $p_3<0$ as the expression as in the above bracket is always positive, wherefore the corresponding system becomes unstable in this limit. For the general case of $r \ne p$ and $X=1=Y$, the above Eqn.(\ref{rnep}) reads as
\begin{equation} p_3=\frac{1}{E^2}\bigg(-e^{p}r(1+p)+e^{-r} r(r-1)\bigg) \end{equation}
After making surface plot of $p_3=0$ with $X=Y=1$, from figure \ref{p3}, we see that there is a sharp value for the Reggeon and Pomeron densities that are found exactly at their values as predicted by GRV model \cite{20b,21b}. Namely, under fluctuations of $\{r,p,E\}$, we find that the optimal values of Reggeon and Pomeron parton densities about the values of $r=0.4$ and $p=0.0245$ for a range of energy scales from $1-2000$ GeV, see fig.(\ref{p3fig}) for a pictorial view. 

This describes fluctuations of the Reggeon-Pomeron based soft interactions at varying initial energy scales from the string fragmentation scale to that of soft interactions and other experimental data such as CDF and UA5 collaborations. Notice that hard interactions happens at domain of a low transfer momentum, where we have a breakdown of the perturbative QCD. Physically, we find that the hard interactions are mostly point-like interactions in their nature, however the soft interaction involves large transverse momentum interaction area. The Pomeronic view of instability is shown in fig.(\ref{p3}) and the corresponding energy view is shown in fig.(\ref{p3energyview}). 

In practice, the partons under soft interactions have an interaction area that satisfy uncertainty relation 
\begin{equation}
\Delta p \Delta b \sim 1,
\end{equation} where $\Delta b$ is the uncertainty in the impact parameter of the partons. On the other hand, the high transverse momentum events are localize in a small region of collisions. Therefore, they maybe approximated by a $\delta$ function. Physically, as discussed in the next section, an energy dependent transverse region yields the soft interactions having $s=1$ with $X=1=Y$ and $r \ne p$, whereby we have the following hypersurface minor
\begin{equation}
p_3=E^{-p-2}\bigg[p(p+1)+\frac{r+p}{\ln E}(1-p\ln E)\bigg]+r(r-1)E^{r-2}
\end{equation}			
Moreover, our analysis is well-suited for the studying of CDF experiments, and P238, UA5, ZEUS collaborations. In this concern, we have provided a case by case detailed analysis of Bjorken scaling and QCD cutoff towards the understanding of the scattering models that of optimally estimated through minijet productions. Notice also that diffraction dissociated particles suffered through incerta phase space decays that are mediated via the system fluctuations if the mass of the excited particle are very low. This is because for higher mass excitations, we have quark-diquark or quark-antiquark pairs that are connected by color strings which undergo Lund string fragmentation, whereby no extra transverse momentum is created.

Physically, it is worth emphasizing that the excited particle of mass more than $10$ GeV are described through the Pomeronic interactions that arise as decaying states with multiple soft and hard interactions. This is generated by meson-proton interaction at the energy about $0.7$ GeV as the mass difference of incoming and excited particles. Such experiments are well-understood through UA4 collaboration \cite{37b}.
\subsection{Multiple soft interactions in Gaussian limit} 
In this section, we discuss the nature of the foregoing section in the case of the limiting Gaussian sample. Hadron-hadron interactions are define through the soft profile function \cite{SIBYLL}, where the softness is introduced through the variance of the impact parameter $b$. Under fluctuations of the proton-proton collision profile as the model embedding function
\begin{equation} \label{msig}
A:\mathcal{M}_2 \rightarrow \mathbb{R}
\end{equation}
with it's coordinate space assignment as
reading as 
\begin{equation} \label{msige}
A_{pp} = \frac{1}{4 \pi(2x+y)} \exp\bigg[-\frac{b^2}{4(2x+y)}\bigg],    
\end{equation} 
where $x= B_p$ and $y=B_s(s)$ characterize the transverse size of the proton and variance of the impact of a shift in interactions of the protons.
Herewith, from Eqn.(\ref{msig}), the stability analysis of proton-proton collision is carried around the critical points of $A_{pp}$ arising as the roots of the flow equation $ A_x=0$ and $A_y =0$ by jointly evaluating the signs of one of the heat capacities $\{A_{xx}, A_{yy} \}$ and the fluctuation determinant $\Delta: = |H|$ as the first and second principal minors of the below fluctuation matrix 
\begin{equation} \label{hm}
H:= \frac{\partial^2}{\partial X^i \partial X^j}A(x,y)
\end{equation}
for the chosen parameter vector $\vec{X}:= (x, y)$ on the surface $\Sigma $ of parameters $\{ x, y\}$. To study fluctuations of the above limiting Gaussian soft profile function, we need to compute its corresponding flow components. In this case, we see that the $x$ and $y$ flow components can be written as
\begin{equation}
\frac{\partial A_{pp}}{\partial x} = \frac{b^2-4(2x+y)}{8\pi(2x+y)^3}\exp\bigg[-\frac{b^2}{4(2x+y)}\bigg] 
\end{equation} 
\begin{equation}
\frac{\partial A_{pp}}{\partial y} = \frac{b^2-4(2x+y)}{16 \pi(2x+y)^3}\exp\bigg[-\frac{b^2}{4(2x+y)}\bigg]
\end{equation}
Hereby, the equilibrium condition is defined by the below flow equations 
\begin{eqnarray}
\frac{\partial A}{\partial x} = 0= \frac{\partial A}{\partial y} 
\end{eqnarray}
In this case, both the above flow equations reduce to an identical criterion 
\begin{eqnarray}
p^2=4(2x+y)
\end{eqnarray}
Here, the fluctuation capacities defined as the second order pure derivatives of $A$ with respect to the system parameters $\{ x, y \}$ simplify as
\begin{eqnarray}
\frac{\partial ^2A_{pp}}{\partial x^2} &=  & \exp\bigg[\frac{-b^2}{8x+4y}\bigg]\frac{(b^4-16b^2(2x+y)+32(2x+y)^2)}{16\pi(2x+y)^5},  \newline \nonumber \\
\frac{\partial^2A_{pp}}{\partial y^2} & = & \exp\bigg[-\frac{b^2}{8x+4y}\bigg]\frac{(b^4-16b^2(2x+y)+32(2x+y)^2)}{64\pi (2x+y)^5}\end{eqnarray}
The associated cross-correlation between transverse size of the proton and the mass center of energy direction in given by 
\begin{eqnarray}
\frac{\partial^2A_{pp}}{\partial x \partial y} & =& \exp\bigg[-\frac{b^2}{8x+4y}\bigg]\frac{(b^4-16b^2(2x+y)+32(2x+y)^2)}{32\pi (2x+y)^5}
\end{eqnarray}
In the light of the above multiple soft interactions in Gaussian limit, the Hessian matrix of the proton-proton soft profile reading as
\begin{equation}
H= \left[ {\begin{array}{cc}
	\frac{\partial ^2A_{pp}}{\partial x^2} & 
	\frac{\partial^2A_{pp}}{\partial x \partial y} \\
	\frac{\partial^2A_{pp}}{\partial x \partial y}
	& \frac{\partial ^2A_{pp}}{\partial y^2} \\
	\end{array} } \right]
\end{equation}
has the eigenvalues $(\lambda_1, \lambda_2)$ of $M$ that are given by
\begin{equation}
\frac{5\exp(-\frac{b^2}{8x+4y})}{64\pi(2x+y)^5}(b^4-32b^2x+128x^2-16b^2y+128xy+32y^2, 0)
\end{equation}
Herewith, we observe that the above Hessian matrix $H$ has a vanishing determinant, that is, we have $\Delta$ = 0. Therefore, in this case, there are infinitely many solution for $x$ and $y$, physically speaking in this Gaussian case, the hadron-hadron collision profile fluctuations yield an indeterminate configuration. This corresponds to white noise in the system. Hereby, we anticipate that there should be higher order interaction determining the shape of the proton or meson profile function. This analysis is carried out in detail in the next section by incorporating uncertainties in the variance of the impact parameter $b$ as the higher order contributions in the Regge trajectory of the constituent hadrons. Further, we offer determination of the amplitude by translating the impact parameter of hadron-hadron interactions with a non-zero amplitude of variations in an energy dependent soft profile function in the Gaussian limit of an arbitrary ensemble of Reggeons and Pomerons.

Notice also that soft interactions are mediated by a pair of quarks or diquarks whose stability can be determine either by optimizing the gluon density profile in its geometric saturation condition in the light of minijet model in high energy limit \cite{Ourgluonpaper}, or as in the present case, also see \cite{SIBYLL}. Moreover, the system stability can equally examined through the Lund string fragmentation between sea quarks that are used to regularize the singular part of quark or diquark distribution functions as long as the applicability of Lund string fragmentation holds. This remains always true for multiple soft interactions as they involve the valence quarks and fractions of the momentum with a given center of mass energy of the hadron-hadron interactions. Physically, we notice that the multiple soft interactions affect the prediction of the hadronic model at an intermediate energy corresponding to the cases of the inelastic and total scattering cross sections during the proton-proton and proton-meson collisions \cite{SIBYLL}, see fig(4) for a qualitative illustration of the regions of stability in its energy view.

Following the same, we have computed the fluctuation structures, namely, the auto correlation and cross correlation functions underlying the hadron-hadron collision profile functions by using an exponential form factor representing the limiting Gaussian shape profile in the transverse space of the constituent protons and/or mesons. In this case, using the first order estimate for $B_s$ and data fit for $B_p$ as the model parameters that characterize the variance of the impact parameter $b$ and the size of a proton respectively, we see that the auto correlations with respect to model parameters take an identical value, except a prefactor. Similarly, the cross correlation function remains identical to the auto-correlations up to a prefactor, as well. 

Substituting the above values of the fluctuation capacities, we see that the respective Hessian determinant vanishes identically for all values of the model parameters $\{x, y\}$. Notice that the process of a diffraction dissociation happens by the quantum exchange phenomena of the energy among colliding particles. Hereby, their exists a large rapidity gap, namely, our proposed intrinsic fluctuation theory based modeling offers a satisfactory understanding of the diffraction-dissociation physics at the level of phenomenology. The full understanding will be given by the method of scattering cross section in the light of multiple minijet productions that should be fitted with the total inelastic cross section of the concerned diffractive events. In our analysis, we have focused on two channel eikonal model for low mass diffraction dissociations, resulting in the superposition of underline low mass diffractive final states. 

Further modifications arise due to the multiplicity distribution, coherence particle productions, quasi elastic scattering of protons, interactions between valence quark-diquark and quark-antiquark pairs that are connected by a fragmentable color string. As far as diffractive excited states are concerned, there are various hadronic interactions that happen above energies $10$ GeV. Indeed, these states can decay through multiple soft and hard interactions that are generated by $\pi$ meson and proton interactions at the center of mass energy $\Delta M$. For an illustration, see UA4 collaboration \cite{37b}, where a single string decay is mediated at the mass difference $\Delta M= 0.7GeV$ between the corresponding incoming particles and excited states.
\section{Comparison of the results}
In this section, we provide the physical meaning and  discussion of the foregoing section by comparing the associated results as follows. 

Below, the symbols have their meanings as follows: DSP: Diffraction dissociation Pomeron, SI: Soft interaction, MJP: Minijet production, LSF: Lund string fragmentation, CDF: CDF Collaboration \cite{CDF}, P238: P238 Collaboration \cite{P238}, UA5: UA5 Collaboration \cite{UA5}, ZEUS: ZEUS measurement collider data for positron and proton \cite{ZEUS} and HI: Hard interactions.
\subsection{Lund string fragmentation}
Below, we give a qualitative description of the Lund string fragmentation of quark-antiquark and diquark-antidiquark pairs that are connected by QCD strings with the respective probability distribution of quark and diquark as in Eqn.(\ref{qdo}) with the model parameters set to their values $\alpha=3.0$ and $\mu=0.35$ GeV as in SIBYLL 2.1 \cite{SIBYLL}. In doing so, we take the fractional mass energy as $x=1/2$, and compute the respective fluctuation quantities at varied energy scales from $1-1800$ GeV as in the Table 1 below. 
\begin{table}
	\begin{center}
		\begin{tabular}{ |c|c|c|c|c|c| } 
			\hline
			E & 1(LSF) & 10(DD) & 30(MJP) & 53(UA5) & 100(HI)\\ 
			$f$ & 0.17675 & 0.17656 & 0.17675 & 0.17677 & 0.17677 \\
			$\frac{f_\alpha}{\log2}$ & -0.16000 &  -0.17656 &  -0.17675 &  -0.17678 &  -0.17677  \\ 
			$f_{\mu}$ & -0.07517 & -0.00123 & -0.00014 & -4.4043 $\times 10^{-5}$ & -1.2374 $\times$ \\
			$\frac{ 10^{-5}f_{\alpha\alpha}}{(\log2)^2}$ & 0.16000 & 0.17656 & 0.17675 & 0.17677 & 0.17677 \\
			$f_{\mu\mu}$ & -0.03820 & -0.00347 & 0.00039 & -8.83843 $\times 10^{-6}$ & -3.5349 $\times 10^-{5}$ \\
			$\frac{f_\alpha\mu}{log2}$ & 0.07517 & 0.00123 & 0.00014 & 4.4043 $\times10^{-5}$ & 1.23736 $\times 10^{-5}$ \\
			$\frac{\Delta}{(log2)^2}$ & -0.01176 & -6.1438 $\times 10^{-4}$ & -6.93122 $\times 10^{-5}$ & -2.22363 $\times 10^-5$ & -6.24893 $\times 10^{-6}$ \\
			\hline
		\end{tabular}
	\end{center}
	\begin{center}
		\begin{tabular}{ |c|c|c|c|c|c| } 
			\hline
			E & 200(UAS) & 210(ZEUS) & 630(P238) & 1000(SI) & 1800(CDF)\\ 
			$f$ & .17678 & 0.17678 & 0.17678 & 0.17677 & 0.17678 \\
			$\frac{f_\alpha}{\log2}$ & -0.17678 &  -0.17678 &  -0.17678 &  -0.17678 &  -0.17678  \\ 
			$f_{\mu}$ & -3.0935 $\times 10^{-6}$ & -2.8059 $\times 10^{-6}$ & -3.1178 $\times 10^{-7}$ & -1.237 $\times 10^{-7}$ & -3.8193 $\times 10^{-8}$ \\
			$\frac{f_{\alpha \alpha}}{(\log2)^2}$ & 0.17678 & 0.17678 & 0.17678 & 0.17678 & 0.17678 \\
			$f_{\mu \mu}$ & -8.83843 $\times 10^{-6}$ & -8.01675 $\times 10^{-6}$ & -8.90783 $ \times 10^{-7}$ & -3.5355 $\times 10^{-7}$ & -1.09121 $\times 10^{-7}$ \\
			$\frac{f_{\alpha \mu}}{log2}$ & 3.0935 $ \times 10^{-6}$ & 2.8059 $\times 10^{-6}$ & 3.11775 $ \times 10^{-7}$ & 1.23744 $ \times 10^{-7}$ & 3.18925 $\times 10^{-8}$ \\
			$\frac{\Delta}{(log2)^2}$ & -1.56243 $ \times 10^{-6}$ & -1.4172 $ \times 10^{-6}$ & -1.57470 $\times 10^{-6}$ & -6.24999 $ \times 10^{-8}$ & -1.92901 $ \times 10^{-8}$ \\
			\hline
		\end{tabular}
		\caption{Lund string fragmentation analysis for quark-antiquark and diquark-antidiquark pairs taken up to five decimal places.}
	\end{center}
\end{table}

Here, we provide numerical values of Lund string fragmentation when the respective distribution function of quark energy is taken as an embedding for the sampling of momentum fractions. In particular, we have provided the nature of fluctuation for various energy scales corresponding to models of soft interactions and their associated experimental counterparts.

From Table(1), we see that Lund string fragmentations happening at energy scale about $1$ GeV correspond to an unstable statistical ensemble under fluctuations of model parameters $\alpha$ and $\mu$.  
Here, the fluctuation capacities have contrasting sign,namely, $A_{\alpha \alpha}$ has a positive sign and the other $A_{\mu \mu}$ has a small negative value. In this case, the fluctuation determinant $\Delta$ takes a negative value of $-0.012$, whereby the respective ensemble undergoes a phase transition in the band of the energy scale corresponding to multiple soft interactions. 

Hitherto, we find a similar observation for a varied range of energies from $1$, GeV to $1800$ GeV that there is always a phase transition and formation of new matter. On the other hand, it is interesting to note that from the determinant of fluctuations, a given phase of matter can be stable if $\Delta$ is positive. This precisely happens when $s>\frac{\mu^2}{x^2}$. The minimum of quark distribution function $A$ requires that $A_{\mu \mu}$ remains positive, that is, we have $s<\frac{3b^2}{2x^2}$. From the above table (1), by taking $x=\frac{1}{2}$, we see that $\Delta$ is positive when $s>4 \mu^2$.

The quark distribution function is minimized if $A_{\mu \mu}$ and $A_{\alpha \alpha}$, is always positive. Therefore, we require that $A_{\mu \mu}$ is positive for $x=\frac{1}{2}$. This happens when $s < 6b^2$. For the given quark mass energy $\mu = 0.35$ GeV, the corresponding determinant takes the value of $0.45$, while the positivity of $A_{\mu \mu}$, requires $s<0.75$ GeV. Hereby, we see that there are five regions in $\alpha-\mu$ plane that are symmetric about the line $\alpha = 0$. For $s< \frac{\mu^2}{x^2}$, we have regions of phase transition where new matter is predicted to be formed. For $s= \frac{\mu^2}{s^2}$, we have a pair of lines of phase transitions. In the case of $s> \frac{\mu^2}{x^2}$, we have a stable region whenever $s<\frac{3b^2}{2s^2}$, whereby the quark distribution function $A$ gets minimized under fluctuations of $\{\alpha, \mu \}$. 

For the energy scales $s \ge \frac{3\mu^2}{2s^2}$, we have unstable region for an increasing value of energies as the quark distribution function gets maximized or it becomes too flat. In this case, when its becomes too flat a separate analysis is required. Such cases are left open for future research. In the case, when the center of mass energy square $s$ is much larger than $\frac{3\mu^2}{2s^2}$, it follows that we have highly excited particles which through the process of fluctuations come down in the stable regions by emitting certain amount of energy. 
\subsection{Lund string fragmentation in Gaussian limit for $u$ and $d$ quarks at $p_{0}= 0.3$}
In this case, we consider model parameter as $a = 0.5$, and $b = 0.8$ as in \cite{SIBYLL} and probability $z= 0.5$, with transverse mass \begin{equation} M_T= \sqrt{m^2+ p_T^2},\end{equation} where the transverse momentum $p_T$ is taken as its mean value $<p_T>$ as in SIBYLL $2.1$ is prescribed \cite{SIBYLL} as

\begin{equation} \label{pt}
<p_T>= \bigg(p_0+0.8 \log_{10}\big(\frac{\sqrt{s}}{30GeV}\big)\bigg)\frac{GeV}{c}.
\end{equation}

In the sequel, we provide observation for the local stability of the quark-antiquark and diquark-antidiquark pairs under the fragmentations of Lund strings in their Gaussian limit and mean energy approximation with its probability distribution as in Eqn.(\ref{qd}). In this case, from Table (2), we observe that the fluctuation capacities always take a positive values as we vary energy from $1$ GeV to $1800$ GeV. Therefore, we notice that the ensemble of quark-antiquark and quark-diquark pairs that are attached to randomly chosen strings yield a locally stable model when one of the model parameters $a$ and $b$ is held fixed. Physically, this indicates that the fragmentation of strings continues until the its mass becomes comparable to a threshold mass of quark and diquark pairs up to an error bar of $(1.1\pm0.2)$ GeV. 

\begin{table}
	\begin{center}
		\begin{tabular}{ |c|c|c||c|c|c| }
			\hline
			s & 1(LSF) & 10(DD) & 30(MJP) & 53(UA5) & 100(HI)   \\
			
			f & 0.08979 & 0.08483 & 0.08197 & 0.08038 & 0.07852 \\  
			$f_a$ & -0.06224 & -0.05880 & -0.05682 & -0.55713 & -0.5443 \\
			$f_b$ & -0.30941 & -0.29835 & -0.29181 & -0.28811 & -0.28375 \\
			$f_{aa}$ & 0.04314 & 0.04076 & 0.03938 & 0.03862 & 0.3773 \\
			$f_{bb}$ & 1.06628 & 1.4934 & 1.03883 & 1.03273 & 1.02540   \\ 
			$f_{ab}$ & 0.21447 & 0.20680 & 0.20226 & 0.19970 & 0.19668 \\
			\hline 
		\end{tabular}
	\end{center}
	\begin{center}
		\begin{tabular}{ |c|c|c||c|c|c| }
			\hline
			s & 200(UA5) & 210(ZEUS) & 630(P238) & 1000(SI) & 1800(CDF)\\ 
			f & 0.07641 & 0.07626 & 0.7274 & 0.07121 & 0.06923 \\  
			$f_a$ & -0.05296 & -0.05286 & -0.05042 & -0.04936 & -0.47984 \\
			$f_b$ & -0.27872 & -0.27836 & -0.26981 & -0.26603 & -0.26107 \\
			$f_{aa}$ & 0.03671 & 0.3664 & 0.03945 & 0.03421 & 0.03326 \\
			$f_{bb}$ & 1.01672 & 1.016019 & 1.0082 & 0.99385 & 0.98543  \\ 
			$f_{ab}$ & 0.19320 & 0.19294 & 0.18702 & 0.18440 & 0.18096 \\ 
			\hline 
		\end{tabular}
		\caption{Lund string fragmentation of Gaussian limit for $u,d$ quark with $p_0= 0.3$ taken up to five decimal places.}
	\end{center}
\end{table}	

Further, we have notice that the model cross-correlation remains always positive at all energy scales from $1$ GeV to $1800$ GeV. In this case, it is worth mentioning that there is no criteria to comment above the global stability of the ensemble when both the parameters $a$ and $b$ are allowed to fluctuate. Such an analysis is left open for future research.

\subsection{Lund string fragmentation in Gaussian limit for $s$ quarks at $p_{0}= 0.45$}
With the parameter values of $a = 0.5$, and $b = 0.8$ and probability $z= 0.5$ and transverse momentum $p_T$ as in Eqn.(\ref{pt}), the respective fluctuation quantities of Lund string fragmentation for $s$ quarks having $p_{0}= 0.45$ are summarized in the Gaussian limit as below.

\begin{table}
	\begin{center}
		\begin{tabular}{ |c|c|c||c|c|c| }
			\hline
			s & 1(LSF) & 10(DD) & 30(MJP) & 53(UA5) & 100(HI)\\  
			f & 0.07937 & 0.07217 & 0.06847 & 0.066502 & 0.06428 \\  
			$f_a$ & -0.055017 & -0.05002 & -0.04746 & -0.04610 & -0.04456 \\
			$f_b$ & -0.28576 & -0.26840 & -0.25914 & -0.25413 & -0.24838 \\
			$f_{aa}$ & 0.038135 & 0.03467 & 0.03289 & 0.03195 & 0.03089 \\
			$f_{bb}$ & 1.02879 & 0.99823 & 0.98085 & 0.97112 & 0.95967  \\ 
			$f_{ab}$ & 0.19807 & 0.18604 & 0.17962 & 0.17615 & 0.17216\\
			\hline 
		\end{tabular}
	\end{center}
	\begin{center}
		\begin{tabular}{ |c|c|c||c|c|c| }
			\hline
			s & 200(UA5) & 210(ZEUS) & 630(P238) & 1000(SI) & 1800(CDF)\\  
			f & 0.61835 & 0.06166 & 0.05775 & 0.05610 & 0.05401 \\  
			$f_a$ & -0.04286 & -0.04274 & -0.04003 & -0.03889 & -0.03744 \\
			$f_b$ & -0.24192 & -0.24146 & -0.23087 & -0.22631 & -0.22043 \\
			$f_{aa}$ & 0.02971 & 0.02963 & 0.027746 & 0.2695 & 0.025948 \\
			$f_{bb}$ & 0.94646 & 0.94550 & 0.92296 & 0.91292 & 0.89969  \\ 
			$f_{ab}$ & 0.16768 & 0.16737 & 0.16002 & 0.15687 & 0.15280 \\ 
			\hline 
		\end{tabular}
		\caption{Lund string fragmentation of Gaussian limit for $s$ quark with $p_0= 0.045$ taken up to five decimal places.}
	\end{center}	
\end{table}

Following the values of the parameters as discussed in Table(2), we find that Lund string fragmentation model of $s$ quarks when it is consider in the Gaussian limit has an identical set of statistical prediction under fluctuations of the model parameters $a$ and $b$ for the strange quark matters. In particular, we see that the fluctuation capacities $f_{aa}$ and $f_{bb}$ as well as the local cross correlation $f_{ab}$ always remain positive. Therefore, our analysis anticipates that there always exists a locally stable strange matter whenever one of parameters is held fixed. Since the fluctuation determinant vanishes, therefore, in a generic setting, it follows that globally new strange quark matters are expected to be formed as the final hadronics states.

\subsection{Lund string fragmentation in Gaussian limit for quark-antiquark pairs at $p_{0}= 0.60$} 
In the sequel, we continue with the same parameter values as $a = 0.5$, $b = 0.8$ and probability $z= 0.5$ with the transverse momentum $p_T$ of the constituent hadrons as in Eqn.(\ref{pt}). In the Gaussian limit, the respective fluctuation quantities of Lund string fragmentation pertaining to quark-antiquark pairs having an initial value of $p_{0}= 0.6$ are given as below.
\begin{table}
	\begin{center}
		\begin{tabular}{ |c|c|c||c|c|c| }
			\hline
			s & 1(LSF) & 10(DD) & 30(MJP) & 53(UA5) & 100(HI)\\ 
			f & .06530 & 0.05713 & .05321 & .05120 & 0.04897 \\  
			$f_a$ & -0.04526 & -0.3960 & -0.3690 & -0.035489 & -0.033944 \\
			$f_b$ & -0.25100 & -0.022916 & -0.21818 & -0.21240 & -0.20587 \\
			$f_{aa}$ & 0.03137 & 0.02745 & 0.02557 & 0.02460 & 0.02353 \\
			$f_{bb}$ & 0.96494 & 0.91921 & 0.89453 & 0.88104 & 0.86544  \\ 
			$f_{ab}$ & 0.17389 & 0.15884 & 0.15123 & 0.14722 & 0.14270 \\
			\hline 
		\end{tabular}
	\end{center}
	
	\begin{center}
		\begin{tabular}{ |c|c|c||c|c|c| }
			\hline
			s & 200(UA5) & 210(ZEUS) & 630(P238) & 1000(SI) & 1800(CDF)\\  
			f & 0.04656 & 0.07626 & 0.04266 & 0.041127 & 0.03921 \\  
			$f_a$ & -0.03228 & -0.032160 & -0.02957 & -0.02581 & -0.02718  \\
			$f_b$ & -0.27872 & -0.19820 & -0.18670 & -0.18187 & -0.17572 \\
			$f_{aa}$ & 0.02237 & 0.02230 & 0.02050 & 0.01976 & 0.01884 \\
			$f_{bb}$ & 0.84775 & 0.84648 & 0.81706 & 0.80423 & 0.78754  \\ 
			$f_{ab}$ & 0.13772 & 0.13736 & 0.12941 & 0.12606 & 0.12180 \\ 
			\hline 
		\end{tabular}
		\caption{Lund string fragmentation in the Gaussian limit for diquark-antidiquark pairs with $p_0= 0.6$ taken up to five decimal places.}
	\end{center}
\end{table}

In this case, we take values of model parameters as discussed above in Table (2) and the  highlight the outcome of the matter fluctuation with diquark. As shown in Table (4), the fluctuation capacities $f_{aa}$, $f_{bb}$, and the local cross correlation $f_{ab}$ all take their positive values in the energy range $1$ GeV to $1800$ Gev. 

Herewith, for various ensembles of $u$, $d$ and $s$ quark-antiquark pairs and diquark-antidiquark pairs, we find that all the limiting ensembles with either $u$ or $d$ quarks or $s$ quarks or diquarks, the behavior of Lund string fragmentation in their Gaussian limits remain identical for various energy values of hadron-hadron interactions mediated via soft interactions.
\subsection{Reggeon-Pomeron model of multiple soft interactions}
To illustrate the notion of soft interaction under Reggeon and Pomeron exchanges, we fix the value of model parameters as $X=1=Y$ describing an equal contribution of the Pomerons and Reggeons respectively. Following the GRV predictions \cite{20,21b} with parton densities as $r=0.4$ and $p=0.0245$, we provide the values of fluctuation quantities in Table(5). In this case, the model embedding function is taken as the energy dependent Reggeon Pomeron soft scattering cross-section for $pp$ and $p \overline{p}$ interactions. 

In particular, we see that there are various behavior of soft interactions that certain systems are stable certain systems are unstable and certain systems are ill-defined depending upon their given energy scale. Below, we discussed behavior of specific model of hadronic interactions and their associated experimental observations.
\begin{table}
	\begin{center}
		\begin{tabular}{ |c|c|c||c|c|c| }
			\hline
			s & 1(LSF) & 10(DD) & 30(MJP) & 53(UA5) & 100(HI)\\  
			f & 2 & 1.27793 & 1.24716 & 1.25650 & 1.27826 \\  
			$f_p$ & 0 & 5.15520 & 8.03602 & 9.64588 & 11.54186 \\
			$f_e$ & 0.3755 & 0.03597 & -0.00262 & 0.01306 & -0.02065 \\
			$f_{r}$ & 0 & -0.72987 & -0.44768 & -0.33146 & -0.23135 \\
			$f_{EE}$ & -0.21490 & 0.00994 & 0.01386 & 0.02047 & 0.025426  \\ 
			$f_{PE}$ & -1 & -1.24574 & -1.37823 & -1.45108 & -1.53592 \\ 
			$f_{Er}$ & 1 & -0.13346 & -0.11326 & -0.090842 & -0.06742 \\ 
			$f_{rr}$ & 0 & 3.36118 & 3.04528 & 2.63120 & 2.13084 \\  
			$f_{pp}$ & 0 & 23.74058 & 54.66415 & 76.59394 & 106.30443 \\
			$\Delta$ & ND & -0.12477 & 0.072150 & 0.05710 & -0.04536 \\
			\hline 
		\end{tabular}
	\end{center}
	\begin{center}
		\begin{tabular}{ |c|c|c|c|c|c| } 
			\hline
			E & 200(UA5) & 210(ZEUS) & 630(P238) & 1000(SI) & 1800(CDF)\\ 
			$f$ & 1.31086 & 1.31341 & 1.37717 & 1.40679 & 1.44629 \\
			$f_P$ & 13.73784 &  13.89753 &  17.167946 &  19.38059 &  21.64419  \\ 
			$f_e$ & -0.02600 & -0.02630 & -0.031300 & -0.032777 & -0.03438 \\
			$f_r$ & -0.15288 & -0.14838 & -0.074773 & -0.055000& -0.037292 \\
			$f_{EE}$ & 0.029078 & 0.029289 & 0.033040 & 0.03426 & 0.03564 \\
			$f_{PE}$ & -1.63301 & -1.64003 & -1.80456 & -1.87764 & -1.97419 \\
			$f_{Er}$ & -0.04672 & -0.04548 & -0.023948 & -0.018019 & -0.012429 \\
			$f_{PP}$ & 145.57487 & 148.62319 & 227.91375 & 267.75271 & 324.4699 \\
			$f_{rr}$ & 1.62000 & 1.58680 & 0.95748 & 0.75986 & 0.55905 \\
			$\Delta$ & -0.03634 & -0.03581 & -0.02638 & -0.02344 & -0.02026 \\
			\hline	
		\end{tabular}
		\caption{Soft interactions mediated by Pomeron and Reggeon exchanges at varied energy scales taken up to five decimal places.}
	\end{center}
\end{table} 

First of all, for energy scale $s=1$ corresponding to Lund string fragmentation, we see that the system is highly ill-defined as first two principal minors of Hessian matrix vanish identically and the third becomes undefined. Nonetheless, for a given pair of parton density of the Reggeons and Pomerons, when the energy scale is varied, we find that the scattering cross section shows maxima with respect to the initial energy $E$.

On the other hand, we find that there are volume instabilities on the fluctuation manifold, that is, the third principal minor becomes negative. This happens for the systems such as diffraction dissociation Pomerons happening at energy scale of $10$ GeV, soft interactions that happen at energy scale of $1000$ GeV and hard interaction happening at the energy scale of $100$ GeV. However, for the case of minijet productions happening at $30$ GeV energy scale, we find that the corresponding ensemble is highly stable as all the associated principal minors remain positive definite. Further, the minijet productions shows a stable behavior even if we fix one or more parameters.

We have further observed that for the multiple soft interactions, namely, their statistical ensembles show both stable and unstable behaviors. In particular, for UA5 collaboration \cite{46,47,48b} conducted at energy scale $200$ GeV, ZEUS collaboration whose data taken at the energy scale of $210$ GeV, P238 collaboration corresponding to at energy scale $630$ GeV and CDF collaboration observed at energy scale $1800$ GeV, we notice that there are volume instabilities, that is, at these scales of energy, there are high chances of the production of new particles. In all of these cases, if we fix one of the fluctuation parameters, then the system becomes stable. On the other hand, for the case of UA5 collaboration, when the data is taken at $53$ GeV, we find that the fluctuating configuration is highly stable, that is, the respective statistical ensemble supports stability of charged particles. 

Physically, in the light of SIBYLL$2.1$, this happens mostly in the central region of soft interactions as described by minijet productions. Note that here we have taken two UA5 experiment \cite{UA5} one at $53$ GeV and other at $200$ GeV that are statistically highly different for a small value of rapidity \cite{SIBYLL}. It is worth mentioning that these two experimental events becomes eventually similar for the events with a large rapidity. However, such observations are not within the scope of present detector of collider measurements. 

Hereby, our analysis clearly distinguishes both the stable and unstable cases within a given error-band. Towards the expansion of our model, further interest include the understanding of Kaons and their decay processes, double diffraction dissociations towards the improvement of SIBYLL$2.1$ and low multiplicity event as triggered by UA5 conditions as mentioned in \cite{UA5}. On the other hand, in all cases we have observed that their are mixed nature of local cross correlations among the model parameters. The above behavior of Reggeon-Pomeron interactions at different energy scales is summarized by the signature of the underlying principal minors as depicted in table (6).	
\begin{table}
	\begin{center}
		\begin{tabular}{ |c|c|c||c| }
			\hline
			SN & $p_1$ & $p_2$ & $p_3$ \\ 
			1 & $+$ & $+$ & $+$  \\  
			$2$ & $+$ & $+$ & $-$  \\
			$3$ & $+$ & $-$ & $+$ \\
			$4$ & $-$ & $+$ & $+$  \\
			$5$ & $-$ & $-$ & $+$   \\ 
			$6$ & $-$ & $+$ & $-$  \\
			$7$ & $+$ & $-$ & $-$  \\
			$8$ & $-$ & $-$ & $-$  \\
			\hline 
		\end{tabular}
		\caption{Minors of the hessian matrix of the soft interactions mediated by Pomeron and Reggeon exchanges at varied energy scales.}
	\end{center}
\end{table}

From the above table (6), we observe the stability of Reggeon-Pomeron system is as follows. In the first case, when all the minors $p_1,p_2$ and $p_3$ are positive the system remains stable and there are no production of new particles. In the case of $2$, $3$ and $4$ when one of the $p_1, p_2,\ \mbox{and}\ p_3$ is negative there are instabilities due to phase transitions, wherefore new particle are produced. The system is stable if we fix one of the variables. In the case of $5,6,7$, there are linear and surface type of instabilities, whereby new particles are prone to be produced. The case $8$ is highly unstable, and in this case new particles are highly prone to be produced. As far as this analysis is concerned, we have volume instability when the volume minor $p_3$ is negative, whereby it happens to be seen in the cases $2, 5, 6, 8$. On the other hand, in the case when $p_2$ is negative, there is a surface instabilities. This happens in the case of $3, 5, 8$. It is worth mentioning that when $p_1$ is negative, the configuration has a linear instability, which precisely happens in the case of $4, 6, 7, 8$. 

Hereby, we have clarified under what cases new matter will be formed or the fluctuations will lead into a crossover, whereby the state of the matter remains the same, i.e., fluctuations leave the state of the matter unchanged. Process of collisions is treated classically where the quantum numbers of hadrons are not exchanged. Phenomenologically, we proposed an intrinsic fluctuation model to comprehend the diffraction dissociation reactions, soft interactions, minijet productions, Lund string fragmentations and hard interactions happening at a high transverse momentum of hadrons. 

\subsection{Gaussian limit of multiple soft interactions}
In order to provide numerical illustration of our model, we take $B_p=x=1.7\ fm$, and $B_s=y=0.01\ fm$, in the near hard limit that is around $1\%$ of the total size of the proton. In the light of hadron-hadron fluctuations, in the limit of exponential form factor where the soft region is described by a fuzzy area that is represented by an energy dependent Gaussian profile function as depicted in Eqn.(\ref{msige}), we define the size of the impact parameter through uncertainty relation $\Delta b \Delta p \sim 1$, that is, we determine the size of impact parameter $b$ of proton-proton collision as $b \sim \frac{1}{m}\sim \frac{1}{\sqrt s}$, where $m$ is the mass of proton observed at center of mass energy $\sqrt s$. 
\begin{table}
	\begin{center}
		\begin{tabular}{ |c|c|c|c|c|c| } 
			\hline
			s & 1(LSF) & 10(DD) & 30(MJP) & 53(UA5) & 100(HI)\\ 
			$A$ & 0.02170 & 0.02333 & 0.02335 & 0.02335 & 0.02335 \\
			$A_{P}$ & -0.01179 & -0.01367 & -0.01369 & -0.01369 & -0.01369  \\ 
			$A_{s}$ & -0.00590 & -0.00684 & -0.00685 & -0.00685 & -0.00685 \\
			$A_{PP}$ & 0.01278 & 0.01603 & 0.01606 & 0.01606 & 0.01606 \\
			$A_{ss}$ & 0.003195 & 0.00401 & 0.00402 & 0.004016 & 0.004016 \\
			$A_{Ps}$ & 0.006340 & 0.00801 & 0.00803 & 0.008031 & 0.008032 \\
			\hline
		\end{tabular}
	\end{center}
	\begin{center}
		\begin{tabular}{ |c|c|c|c|c|c| } 
			\hline 
			s & 200(UA5) & 210(ZEUS) & 630(P238) & 1000(SI) & 1800(CDF)\\ 
			$A$ & 0.02335 & 0.02335 & 0.02335 & 0.02335 & 0.02335 \\
			$A_{P}$ & -0.01369 &  -0.01369 &  -0.01369 &  -0.01369 &  -0.01369  \\ 
			$A_{s}$ & -0.00685 & -0.00685 & -0.00685 & -0.00685 & -0.00685 \\
			$A_{PP}$ & 0.01606 & 0.01606 & 0.01606 & 0.01606 & 0.01606 \\
			$A_{ss}$ & 0.004016 & 0.004016 & 0.004016 & 0.00402 & 0.004016 \\
			$A_{Ps}$ & 0.008032 & 0.008032 & 0.008032 & 0.00803 & 0.008032 \\
			\hline
		\end{tabular}
		\caption{Gaussian limit of soft interactions mediated by Pomeron and Reggeon fluctuations via the proton-proton collisions at varied energy scales taken up to five decimal places.} 
	\end{center}		
\end{table}

From Table(7), we observe that fluctuations mediated by proton-proton collisions with the corresponding soft profile function as the model embedding that the system is generically ill-defined for various values of the energy scales running from $1$ GeV to $1800$ GeV. This is because of the vanishing of the fluctuation determinant under variations of the variance of the soft interaction region and the transverse size of a proton. However, when one of the parameters is fixed, we find that the system become well-defined and its corresponds to a minimum of the soft profile function of the proton-proton collision at various energy scales from $1$ GeV to $1800$ GeV. 

In all such physically acceptable energy scales ranging from $1$ GeV to $1800$ GeV, we notice that the cross correlation between the center of mass direction and the transverse direction of a proton always possesses a positive valued local correlation.
\section{Extension of the SIBYLL 2.1}
In this section, we provide an extension of the soft profile function for proton-proton and proton-antiproton collisions when there is a certain degree of fuzziness in their soft interaction region. This can be introduced in two ways: the first is realized by introducing a shift in the radius of the fuzzy area that is equivalent to the off-mass-shell condition. The other one is done by introducing certain higher order corrections in the Regge trajectory of the constituent hadrons through the variance of their soft interactions region. This is realized by performing Taylor expansion in $B_s(s)$.
\subsection{Translation of the soft interaction region}  
\begin{center} 
\begin{figure}[t!]
	\vspace{-2.5cm} 	\centering
	\includegraphics[width=14.0cm,angle=0]{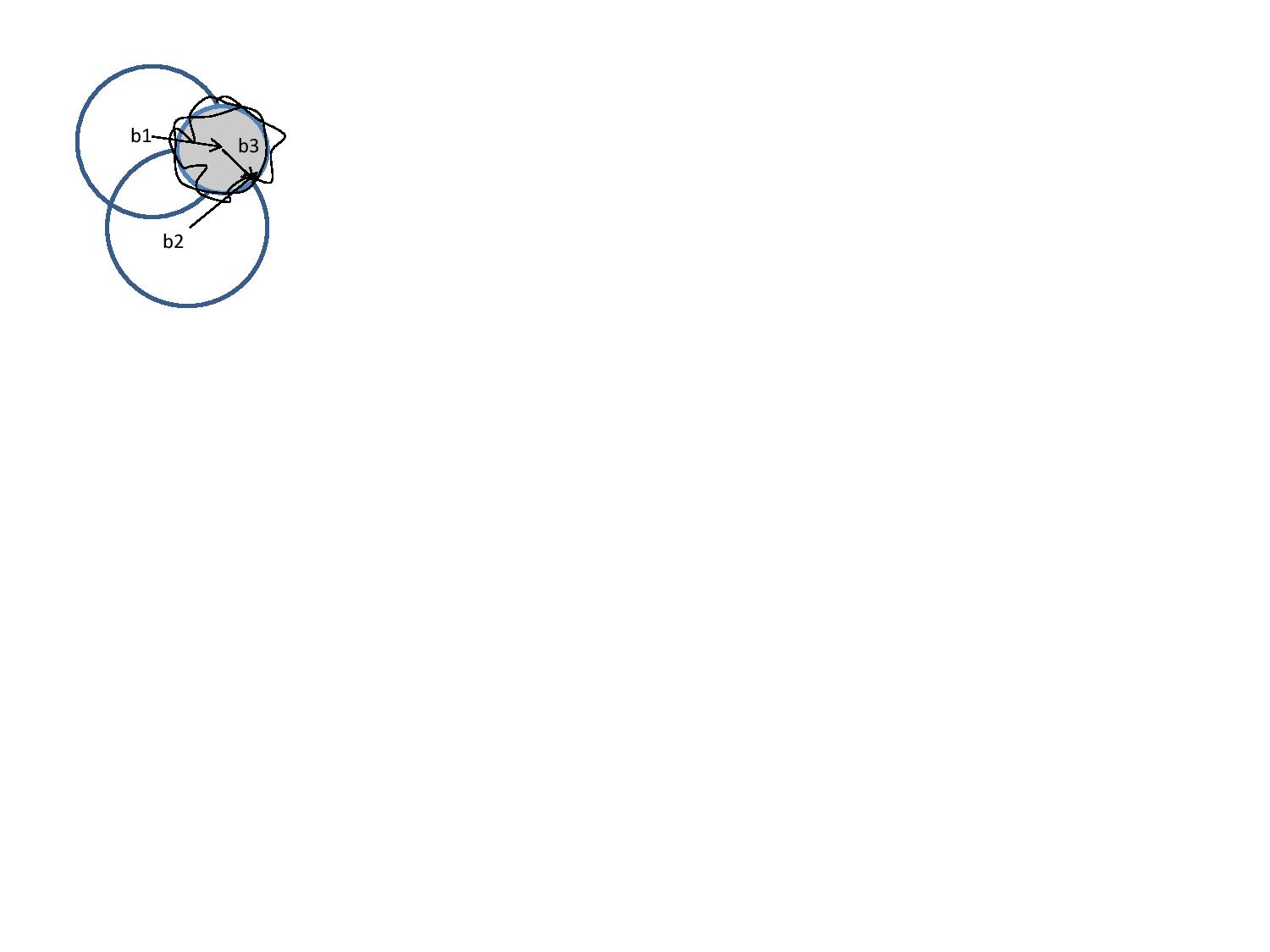}
	\vspace{-6.5cm}
	\caption{The region of the soft interactions under higher order corrections mediating ziggling in boundary of the soft interaction area through the Regge trajectory of the constituent hadrons with uncertainty involving events of a low transverse momentum $p$ and the impact parameter $b$ of the soft region as $\Delta p \Delta b \sim 1$ in contrast to hard interactions \cite{SIBYLL}.} \label{softf}

\end{figure}
\end{center}

In this subsection, we focus on translation of the soft interaction region that physically correspond to the evaluation of soft proton-proton amplitude at an off-mass-shell condition \cite{SIBYLL}. Namely, the soft interaction region has certain degree of uncertainties or fuzziness in the region of interactions as shown in fig.(\ref{softf}) that are parameterized by a Gaussian soft profile function. 

In this case, the region of interactions gets shifted as $b_3 \rightarrow b_3+b_0$, whereby the on-shell condition $b_3=b-b_1+b_2$ as in \cite{SIBYLL} gets modified as 
\begin{eqnarray}
b_3+b_0=b-b_1+b_2
\end{eqnarray}
Thus, we notice that the impact parameter $b$ defining the soft interactions of hadron model as in SIBYLL 2.1 reduces as $b \rightarrow b-b_0$. 
Hereby, the soft profile function as obtained for proton proton collision in the limit of exponential form factor with the transverse size of a proton as fitted by SIBYLL 2.1 data and the variance of the radius of soft region being small, gives the Gaussian profile function with a modified impact parameter $b'=b-b_0$. 

Hence, under fluctuation of the transverse size of a proton $B_p$ and the variance $B_s$ of the soft dimension $b_3$ of hadron-hadron interactions, the underlying stability analysis results into the same conclusion without a translation of the radius of regions of soft interactions.  

In short, our model classifies whether at there will be production of new particles during the fluctuation of transverse momentum $p_T$ and the effective impact parameter $b'$ governing the soft interactions. Systematically, the corresponding region of the soft interactions is shown as below in fig.(\ref{softf}).
\vspace*{0.0cm}	
\subsection{Inclusion of acceleration to hadrons}
In this subsection, we discuss the stability analysis of soft interaction configurations when the constituents hadrons are accelerated or have jerks. At a given centre of mass energy square $s$ of the associated hadrons, this is introduced by considering Taylor series expansion in logarithmic scale \begin{eqnarray} \lambda = \ln(\frac{s}{s_0}) \end{eqnarray} in the Regge trajectory, where $s_0$ is the initial energy scale. For a given soft interactions of radius $b_3$ as illustrated in fig.(\ref{softf}), when $B_s(s)$ is characterized in terms of the Regge trajectory $\alpha(s)$ of an emitted particle, its variance can be expanded as a function of the energy $s$ as below
\begin{eqnarray}
B_s= B_0+ \alpha'k_1+ \alpha^{''}k_2 + \ldots,
\end{eqnarray}
where $k_1=\ln(\frac{s}{s_0})$, and  $k_2 = k_2(s)$.
To evaluate the soft profile function $A^{soft}$, we need to calculate the inverse of $B_s$. With the above expression, it follows that we have
\begin{eqnarray}
\frac{1}{B_s} = \frac{1}{(B_0+ \alpha^{'} k_1)} [1+ \frac{\alpha^{''} k_{2}}{B_0+\alpha^{'}k_1}]^{-1}
\end{eqnarray}
In order to make Taylor expansion of $\frac{1}{B_s}$, define the expansion parameter
\begin{eqnarray}
t = \frac{\alpha^{''}k_{2}}{B_0+\alpha^{'}k_1}
\end{eqnarray}
Thus, for small values of $t$, the inverse of $B_s$ is given by
\begin{eqnarray}
\frac{1}{B_s} = \frac{1}{(B_0+ \alpha^{'} k_1)} (1-t+t^2-t^3+\ldots)
\end{eqnarray}
Up to linear order in $t$, we have $B_s'= B_s(1+t)$, herewith the soft profile function $A^{soft}$ for hadron-hadron interactions \cite{SIBYLL} defined as
\begin{eqnarray}
A^{soft} = \frac{1}{4 \pi B_s} \exp{(-\frac{|b_3|^2}{4B_s})}
\end{eqnarray}
can be expressed as a function of the expansion parameter $t$ as the below expression
\begin{eqnarray}
A^{soft} =\frac{B_0+ \alpha^{'} k_1}{4 \pi} (1-t+t^2- \ldots) \exp\bigg[{-\frac{|b_3|^2}{4(B_0+ \alpha^{'} k_1)}}\bigg](1-t+t^2-\ldots) 
\end{eqnarray}
In the small limit of $t$, we see that the above amplitude simplifies as
\begin{eqnarray}
A^{soft}= \frac{B_0+ \alpha^{'} k_1}{4 \pi} (1-t) \exp\bigg[{-\frac{|b_3|^2}{4(B_0+ \alpha^{'} k_1)}\bigg](1-t)}
\end{eqnarray} 
Herewith, with the fact that $b_3 \rightarrow \sqrt{1-t} \ b_3$, under small $t$ limit, we observe that there is an order by order conservation relation
\begin{eqnarray}
B_s'.B_p' = B_s. B_p,
\end{eqnarray}
where $B_s$ is the variance of $b_3$ and $B_p$ is the transverse size of a proton upto a velocity correction, and $B_s', B_p'$ are that of their respective quantities upto acceleration terms in the Regge trajectory expansion, that is, they describe the formation of a soft profile function for accelerated hadrons.

Given an energy dependent region for hadron-hadron collision, the soft profile function is described as 
\begin{eqnarray}
A^{soft}_{yz} (s,b)=\int d^2b_1 d^2b_2 d^2b_3A_y (b_1)A_z(b_2)A^{soft}(s,b_3) \delta^2{(b_1-b_2+b_3-b)} 
\end{eqnarray}
Substituting the transformed value of $b_3 \rightarrow \sqrt{1-t} \ b_3$, it follows that the above soft profile function satisfies
\begin{eqnarray}
A^{soft}_{yz} (s,b)=\int d^2b_1 d^2b_2 d^2b_3A_y (b_1)A_z(b_2)A^{soft}(s,b_3\sqrt{1-t}) \delta^2{(b_1-b_2+b_3-b)} 
\end{eqnarray}
Assuming the exponential form factor for the incoming and outgoing states of proton or meson profile functions \cite{SIBYLL}, the above soft profile function reads as
\begin{eqnarray}
A^{soft}_{yz} (s,b)= \int d^2{b_1}d^2{b_2}d^2{b_3}exp\bigg[-\frac{|b_1|^2-|b_2|^2}{4B_p}\bigg]\exp\bigg[\frac{-(1-t)|b-b_1-b_2|^2}{4B_s}\bigg]
\end{eqnarray}
In this case, we see hereby that the final answer for the proton-proton collision profile $A^{soft}$ retains the same form with the following redefinitions
\begin{eqnarray}
b_1^{'}&=& {b_1 \sqrt{1-t}}, \nonumber \\ \newline
b_2^{'}&=&{b_2 \sqrt{1-t}}, \nonumber \\ \newline
b^{'}&=&{b\sqrt{1-t}}
\end{eqnarray} 

It is clear that the Gaussian integral when its variance get scaled, the resulting expression involving the shift in the fluctuation variable gets oppositely scaled so that the net Gaussian profile remain the same. Herewith, we conclude that the stability criteria as discussed earlier for proton-proton collision remains physically identical even if the underlying hadrons are accelerated. 

In other words, in small $t$ limit, the physics of fluctuation in the light of soft interaction is independent of the accelerations up to redefinition of the region of interactions and that are of their model parameters. In particular, the nature of statistical interactions of constituent hadrons is invariant under the re-parametrization of model parameters as the variance of $b_3$, and the transverse size of the proton. In short, we observe that such contributions arising due to an acceleration of the constituent hadrons yield fuzziness in the region of soft interactions, see fig.(\ref{softf}) for a geometrical interpretation.

\subsection{Inclusion of jerk to hadrons}
Following the above observation concerning the accelerated hadrons, we include the third order derivative in the variance $B_s(s)$ of $b_3$ as an expansion in the Regge trajectory as a function of the running center of the mass energy square with respect to a given initial energy scale.

In this case, we can equally assume that the second derivative of the Regge trajectory is much larger than its third derivative. Thus, up to third derivative in Regge trajectory, the behavior of hadrons that are accelerated and having jerks is given by the following expansion 
\begin{eqnarray}
B_s =B_0 + \alpha^{'}k_1+ \alpha^{''} k_2 + \alpha^{'''} k_3
\end{eqnarray}
Physically, for the case of Regge trajectories with $\alpha^{''}>> {\alpha}^{'''}$,
there are many experiments running with center of mass  energy $s= 500 GeV^2$. The above expression of $B_s$ can be rewritten as
\begin{eqnarray} B_s = (B_0 + \alpha^{'}k_1+ \alpha^{''} k_2 )
\bigg[1+\frac{\alpha^{'''} k_3}{B_0+ \alpha^{'}k_1+\alpha^{''}{k_2}}\bigg]
\end{eqnarray}
Defining the new expansion parameter as 
\begin{eqnarray} 
t=\frac{\alpha^{'''} k_3}{(B_0+ \alpha^{'}k_1+\alpha^{''}{k_2})}
\end{eqnarray}
In this case, it follows that an inclusion of jerk to the motion of constituent hadrons results into the Gaussian profile with a re-scaling of the variance of $b_3$, and the transverse momentum of a proton as given before as in the case of accelerated hadrons, that is, we have the following transformations
\begin{eqnarray}
B_s^{'}&=& B_s (1+t), \nonumber \\ \newline
B_p^{'}&=&B_p (1-t)
\end{eqnarray} 
Here, $B_s$ and $B_p$ are properties the constituent protons under the collision. In this case, we find that the constituent hadronic scales according as the concerning protons whose soft interactions are given by 
\begin{eqnarray}
b_1^{'}&=& {b_1 \sqrt{1-t}}, \nonumber \\ \newline
b_2^{'}&=& b_2 \sqrt{1-t}, \nonumber \\ \newline
b_3^{'} &=& b_3 \sqrt{1-t}
\end{eqnarray} 
Therefore, we see that the Gaussian profile even upto the jerk contributions remain unstable under translation in the soft interaction region and dilation in the impact parameter. Here, in the case of dilation as the higher order Regge corrections, it follows that our conclusion that the Gaussian soft profile remains ill-defined hold at an arbitrary order of expansions. Namely, for a given impact parameter $b$, at $(n+1)$-th order of the expansion of its variance
\begin{eqnarray}
B_s =B_0 + \alpha^{'}k_1+ \alpha^{''} k_2 + \ldots + \alpha^{n+1} k_{n+1},
\end{eqnarray} 
the underlying ensemble corresponds to an ill-defined statistical basis. This represents the case of certain white noise in the system. Hereby, we see that, under higher order corrections in the Regge trajectory of the constituent hadrons, the ensemble possesses the same statistical properties as that of the original system up to a scaling of the impact parameter $t$ of their soft interactions as the following expansion parameter
\begin{eqnarray}
t= \frac{\alpha^{n+1} k_{n+1}}{\sum_{i=0}^n \alpha^{i}{k_i}},
\end{eqnarray} 
where $\alpha^{n}$ denotes the $n$-th derivative of $\alpha(s)$ and $k_0=B_0$ is the initial value of the variance of the impact parameter of soft interactions. This offers a modified contribution to $B_s(s)$ in the logarithmic scale as a function of the center of mass energy square with reference to an initial energy scale. The concerning jerks to the hadrons under collision introduce certain bumps in the soft region of interactions. Namely the boundary of soft interaction area gets fluctuations as shown in fig.(\ref{softf}). The analysis for the case of a large $t$ is left open for future research and developments.
\section{Conclusion} 
In this paper, we have studied hadronic soft fluctuations and provided criteria for the stability or the formation of new particles using various low energy phenomenological model of QCD and their applications to cosmic ray event generators. Our particular focus is given on an intrinsic analysis towards the designing of accelerators beyond the present energy scales as in SIBYLL 2.1. In this concern, we have examined stability of low energy effective field theory configurations composed of quarks and diquarks and multi particle soft interactions, incorporating transverse resolution of hard interactions that are comparable to proton size where the nonlinear effects and strong coupling situation cannot be ignored. In the sequel, we have considered Reggeon-Pomeron fluctuations involving multiple soft interactions. Following the Regge theory model of hadrons, we have given an explicit illustration of the wall of stability/instability under fluctuations of Reggeon-Pomeron density and initial energy scale as the model parameters.

In the above setting, we have shown that the numerical value of the Reggeon Pomeron densities as predicted by Donnachi and Landshoff \cite{DL} arise as the optimal value of our model. For proton-proton and proton-antiproton interactions, the optimal parton density identically matches with the GRV parton densities of Reggeon and Pomeron \cite{20,21b}. The wall of stability/instability physically describes the nature of soft interactions that are transformation of hard interaction domain with a low transfer momentum. This is well justified in the real mass non perturbative QCD, where the concerning soft interactions happens at a large transverse interaction area. Due to fluctuation of Reggeon and Pomeron densities, the amplitude of soft interaction gets lower whereby we have low transverse momentum partons that interact in a spreaded region. Moreover, we have discussed soft interactions whose profile function in long time effect simplifies as a Gaussian function. Hereby, in the limit of small variance of the region of soft interactions and by using exponential form factor for proton profile function, we have provided stability criteria for an ensemble of hadrons. Furthermore, we have provided qualitative discussions and future directions towards the study of random observation samples of multiple soft interactions.

Notice also that the corresponding hard interactions arise without the variance of impact parameter in contrast to soft interactions. In this perspective, we have found that hadron hadron profile fluctuations in the limit of Gaussian approximation yield an indeterminate configuration. Therefore we anticipate that there should be higher order interactions in optimally determining the shape of the proton or meson profile function. This analysis is equally carried out in detail under the translation of the soft interaction region and expansion in the variance of radius of soft interactions as the running mass energy square of partons is varied.In both of the above cases, we have observed that the translation in the region of soft interactions that could physically be considered as off-mass-shell-condition. We equally consider the expansion over the variance of the radius of soft interactions as a function of the running mass energy square as higher order corrections in the Regge trajectory of hadrons. Hereby, the nature of hadron-hadron profile fluctuations is discussed where the centre of mass energy and the transverse size of a proton remains intact. In addition, even after incorporating the higher order corrections in Regge trajectory of the constituent hadrons, or going far away from the on-shell condition, our analysis predicts a production of new particles in the light of multiple soft interactions as introduced in the minijet model that is compared with its experimental incarnations such as the UA5  collaborations \cite{46,47,48b} and references therein. 

As far as an ensemble of quarks and diquarks are concerned, we have provided the critical point analysis of the Lund string fragmentation and classified the regions where we have an unstable/stable statistical basis under variations of the effective quark mass and fractionation index. It is shown that new particles are prone to be produced in the region of $|\mu|\le \sqrt{s}/2$, where $s$ is the square of the centre of mass energy of hadron-hadron interactions. Hereby, if an ensemble of quarks that are connected by the color strings is stable, then the respective ensemble of diquarks becomes unstable. This is due to energy conservation of quarks and diquarks whose distributions when summed up gives unity. Subsequently, we determine the optimal region of Lund string fragmentation under fluctuation of quark-antiquark and diquark-antidiquark pairs that are randomly chosen to be connected by color strings. Here, the combinations happen in such a way that we have conservation of both the flavour and color degree of freedoms. In practice, this is performed for a given primodial transverse momentum pair of quark-antiquark and diquark-antidiquark in the Gaussian approximation. Under fluctuations of the model parameters at a fixed transverse mass of the parent quarks and diquarks, we have found that fluctuations continue in a damped manner until the remaining mass of the string combining quark-antiquark or diquark-antidiquark becomes smaller than a threshold mass. 

Physically at this point strings start fragmenting whereby two final hadrons are produced. Our analysis doesn't stop here, but it continue for soft fluctuations of hadron-hadron profile functions. Here, the radius of soft interactions and the corresponding transverse momentum are constrained through an energy dependent uncertainty relation. In this case, we have shown that there are possibility of local stabilities as far as hadronic interactions are concerned. We have equally tabulated the nature of soft interactions for various fluctuating configuration corresponding to SIBYLL 2.1. Our analysis is well-suited for multiple soft interactions involving the valence quarks and diquarks satisfying Lund string fragmentation. Hereby, we have studied collider physics in the framework of perturbative QCD and beyond. In this case, we have used minijet model with multiple soft interactions, QCD improved parton model, hadron-hadron collisions and strong interaction models whose coupling constant satisfying a geometric saturation condition \cite{23}.8 As far as Reggeon Pomeron ensembles are concerned, we have discussed their stability structures at various values of the model parameters.  

For an identical linear combination of Reggeon and Pomeron contribution to soft scattering cross section, some interesting cases include $r=p=1$, $p=-1$ and $r=1$ and others $p=0$ and $r$ satisfying a non linear equation, $r=0$ and $p=0$, $r=0$ and the running mass energy as $e$ times the initial mass energy, and $p=-r$ that has solution as $r=0=p$ and $r=1$ and $p=-1$. In all such cases, our model has provided an instructive classification scheme concerning the phase transitions and the stability of an experimental ensemble composed of nucleons and/or hadrons. Further extensions of our model include analysis of semi-superposition model and the full-Glauber model. Understanding of the antiproton and its behaviour in the central region of quark-diquark fragmentation is anticipated to be examined further in the light of E735 collaboration \cite{56} and its recent incarnations. Further, a treatment of (in)coherent diffraction dissociation collisions is sought towards hadron-nucleus and nucleus-nucleus interactions through the implementation of the full Glauber picture of proton-proton interactions. 

It is interesting equally to extend our model for ensembles involving charm and strange quarks, whereby gluons play important role in understanding the physics and the optimal designing of muonic and neutrino based detectors. Moreover, our proposition finds high importance in cosmic ray observations and analysis of cosmic ray air shower data where the parton density finds a saturation with an energy dependent cutoff with respect to the transverse momentum. Finally,  in the framework of hadron-nucleus and nucleus-nucleus interactions, it is worth understanding the central region whereby the pseudo-rapidity and multiplicity distribution of charge particles. As we have extended our analysis for soft interaction regions over off mass shell condition and higher order correction in the Regge trajectory, it would be equally pertinent to extend the same for a good fit of the rapidity and Feynman $x$ distributions involving fixed targets as in SIBYLL 2.1 and targets with accelerations and jerk as the second and third derivatives of the Regge trajectory of concerned protons or antiprotons as discussed in the foregoing section. For experiments with fixed targets, prospective investigations include the analysis corresponding to $NA49$ collaboration and NAL data \cite{49b}. A similar analysis can be provided as a consequence of our proposition with higher order correction in the Regge trajectory for examining the jet quenching dictated mechanisms and color coherence \cite{131}. Such analysis is left open for future research and developments.
\section*{Acknowledgments}
B.N.T. would like to thank the Yukawa Institute for Theoretical Physics at Kyoto University.
Discussions during the workshop YITP-T-18-04 ``New Frontiers in String Theory 2018" were
useful for the realization this work. Rahul Nigam and Rahul Thakur  would like to thank BITS Pilani Hyderabad Campus to provide all the infrastructure necessary to carry out this work.

\end{document}